\documentclass[prd,superscriptaddress,amsmath,nofootinbib]{revtex4}
\usepackage[dvips,final]{graphicx}
  \usepackage{amssymb,amsmath,epsfig,bm,pifont}
   \graphicspath{{./NewFigs}}
\usepackage[normalem]{ulem}

\newcommand{\Li}[1]{\mathop{\mathrm{Li}_{#1}}}

\newcommand{\A}{{\mathcal{A}}}
\newcommand{\M}{{\mathfrak{A}}}
\newcommand{\I}{{\mathcal{I}}}
\usepackage{relsize}
\newcommand{\babar}{{\mbox{\slshape B\kern-0.1em{\smaller A}\kern-0.1em
            B\kern-0.1em{\smaller A\kern-0.2em R}}}
\def\MSbar{\relax\ifmmode\overline                        
            {\rm MS}\else{$\overline{\rm MS}${ }}\fi}     
           }                                              
\def\MSbar{\relax\ifmmode\overline                        
            {\rm MS}\else{$\overline{\rm MS}${ }}\fi}     
\def\1{\hbox{{1}\kern-.25em\hbox{l}}}

 \date{\today}
\newcommand{\mycomment}[1]{\iffalse  #1  \fi}                               
\def\Im{\relax{\textbf{Im}{}}}                            
\def\Re{\relax{\textbf{Re}{}}}                            
\newcommand{\Ds}{\displaystyle}                           

\newcommand{\be}{\begin{equation}}\newcommand{\ee}{\end{equation}}%
\newcommand{\bd}{\begin{displaymath}}\newcommand{\ed}{\end{displaymath}}
\newcommand{\bit}{\begin{itemize}}                        
 \newcommand{\eit}{\end{itemize}}                         
\newcommand{\ben}{\begin{enumerate}}                      
 \newcommand{\een}{\end{enumerate}}                       
\newcommand{\baa}{\begin{array}{lll}}                     
 \newcommand{\eaa}{\end{array}}                           
\newcommand{\ba}{\begin{eqnarray}}                        
 \newcommand{\ea}{\end{eqnarray}}                         
\newcommand{\la}{\label}                                  
 \newcommand{\nn}{\nonumber}                              
\newcommand{\gev}[1]{\relax\ifmmode{\text{GeV}^{#1}}      
                     \else{GeV$^{#1}${ }}\fi}             
\newcommand{\Gev}{\relax\ifmmode{\text{GeV}}              
                     \else{GeV{ }}\fi}                    
\newcommand{\Mev}{\relax\ifmmode{\text{MeV}}              
                     \else{MeV{ }}\fi}                    
\def\MSbar{\relax\ifmmode\overline                        
            {\rm MS}\else{$\overline{\rm MS}${ }}\fi}     
\def\as{\relax\ifmmode \alpha_s\else{$ \alpha_s${ }}\fi}  
\def\abar{\relax\ifmmode{\bar{a}}\else{$\bar{a}${ }}\fi}  
  \def\ie{\hbox{\it i.e.}{ }} 
   \def\eg{\hbox{\it e.g.}{ }}  

\usepackage{color}

\definecolor{green}{rgb}{0.133,0.56,0}

\definecolor{DarkGreen}{rgb}{0.04,0.5,0.1}
 
\definecolor{GrayW}{rgb}{0.50196,0.50196,0.50196}
 \newcommand{\GrayW}[1]{{\color{GrayW} #1}}

  \newcommand{\BluTn}[1]{\textcolor{blue}{#1}}
   
\begin{document}
\title{Extending the LCSR method to the electromagnetic pion form factor at low momenta
       using QCD renormalization-group summation}

\author{C\'esar Ayala}
\email{cayalan@academicos.uta.cl}
\affiliation{Departamento de Ingenier\'ia y Tecnolog\'ias, Sede La Tirana, Universidad de Tarapac\'a, Av.~La Tirana 4802, Iquique, Chile\\}

\author{S.~V.~Mikhailov}
\email{mikhs@theor.jinr.ru}
\affiliation{Bogoliubov Laboratory of Theoretical Physics, JINR,
             141980 Dubna, Russia\\}

\author{A.~V.~Pimikov}
\email{pimikov@mail.ru}
\affiliation{Bogoliubov Laboratory of Theoretical Physics, JINR,
	         141980 Dubna, Russia\\}

\date{\today}
\begin{abstract}
We obtain the electromagnetic pion form factor (emFF) $F_\pi$ for spacelike mid-range of momentum transfer in QCD.
We use renormalization group (RG) summation within the light cone sum rules (LCSRs) to obtain the QCD radiative corrections
to the  $F_\pi$ and involve contributions of the leading twist 2 and,
 twists 4, 6.
The additional conditions to apply here this RG summation are discussed in details.
The strong coupling constants in this approach are free of Landau singularities,
which allows one to go down to the lower transferred momentum $Q^2$.
The prediction of the calculations performed reproduces the experimental data below/around $Q^2= 1$~GeV$^2$
significantly better than analogous predictions based on a fixed-order power-series expansion in the standard QCD.
\end{abstract}
\pacs{11.10.Hi,12.38Bx,12.38.Cy,12.38.Lg}
\maketitle

\section{Introduction}
\label{sec:intro}
\begin{figure}[h]
\includegraphics[width=0.9\textwidth]{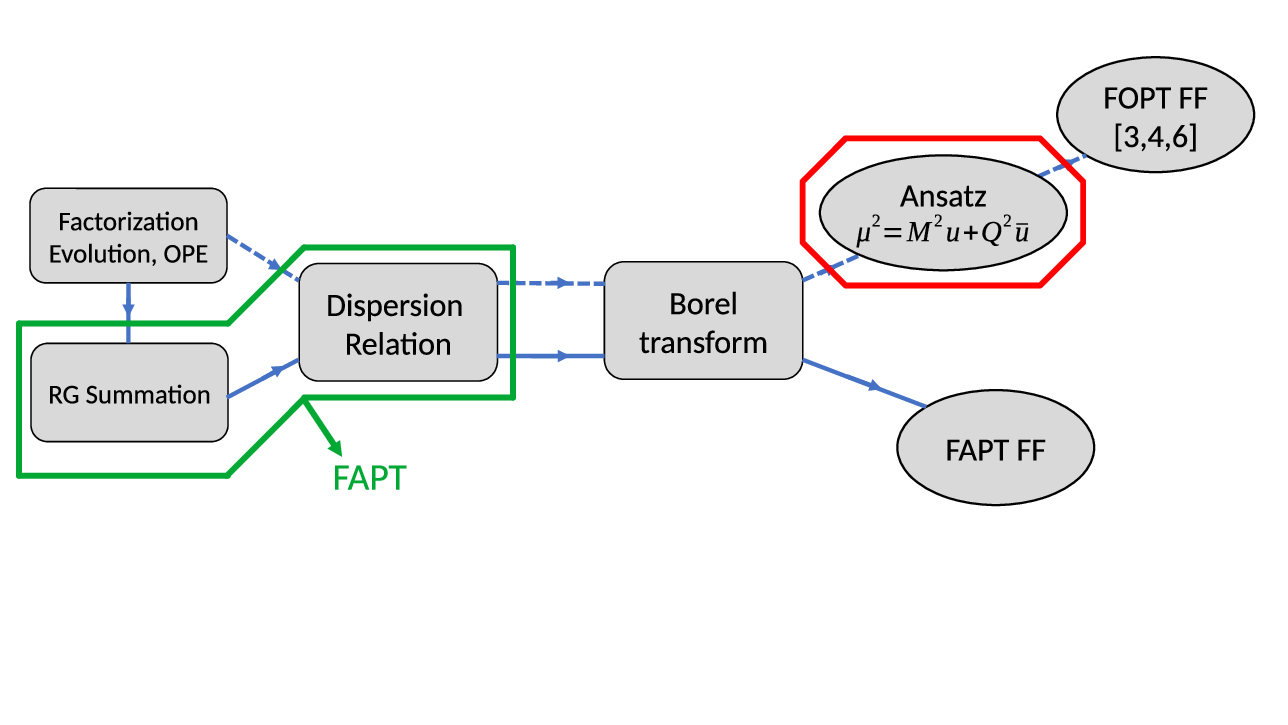}
\vspace*{-30mm}

\caption{\footnotesize{\label{fig:diagram}
The upper dashed line connected blocks show the way of the standard FOPT LCSR,
here $\mu^2=\mu^2_F=\mu^2_R$ is renormalization/factorization scale, $M^2$ is the Borel parameter.
The lower solid line illustrates our approach to the LCSR with the preliminary renormalization group summation.
The crucial elements of our approach are presented in blocks ``RG summation'' and ``Dispersion relation'',
which are taken in a frame.}}
\end{figure}

The hadron form factor (FF) is the appropriate candidate to test  the  QCD in the meaningful range of the
transferred momentum, \eg, \cite{Efremov:1979qk}.
When we delve into the mid- to low-energy range, the properties of analyticity and unitarity in QCD come to the forefront. These properties guide us toward the light cone sum rule (LCSR) and dispersion relations   \cite{Khodjamirian:1997tk,Braun:1999uj,Bijnens:2002mg}, shaping our understanding of this energy range.
Let us delve into the pion electromagnetic FF (emFF), which is essentially a parametrization of the four-vector part of the expectation value of the electromagnetic current
\begin{equation}
    \left\langle\pi^{+}\left(p_2\right)\left|j_{\mu}^{em}\right| \pi^{+}\left(p_1\right)\right\rangle=\left(p_1+p_2\right)_\mu F_\pi\left(q^2\right),\,q= p_1-p_2.
\end{equation}
As we mentioned, the dispersion relation is crucial to extend the analyticity structure of this physical  quantity.
To this end, 
 we will invoke the obtained  knowledge of the pion transition FF (TFF) \cite{Mikhailov:2021znq} because of their analogy, i.e., the replacing axial current $j_{\mu 5}$ with electromagnetic one $j_{\mu}^{em}$ in the corresponding LCSR where the  dominance on the light cone leaves to an expansion in nonlocal operators and then to pion distribution amplitudes (DAs) $\varphi_{\pi}(x)$.
 We will follow the LCSRs as a general frame to obtain the emFF without appealing to an additional ansatz/assumption about the effective scale $\mu^2$ of the QCD coupling constant $a_s(\mu^2)=\alpha_s(\mu^2)/(4\pi)$ after borelization
 (see upper way in Fig.\ref{fig:diagram}).
The $O(\alpha_s)$ corrections to the  LCSR for the emFF within fixed-order perturbation theory (FOPT) were
calculated for the first time in \cite{Braun:1999uj}.
Then, the LCSR was updated from different aspects in \cite{Bijnens:2002mg,Cheng:2020vwr}.
We use the  experimental data  from the JLab collaboration
in \cite{PhysRevC.78.045203} and that cover the intermediate energy range $0.6\leq \left(-q^2=Q^2 \right)\leq 2.45$ GeV$^2$, which is well suited for testing the approach above.
The meaningful analysis of this data processing and the results for the intrinsic pion structure were
presented in \cite{Khodjamirian:2011ub} and \cite{Cheng:2020vwr}.
To be more consistent, we start with the initial object to construct the corresponding  LCSR -- the correlation amplitude $T_{\mu\nu}$,
\begin{subequations}
\ba
T_{\mu\nu}(p,q)\! &=& i\!\int\! d^4 z \,e^{iqz}
\langle 0| T\left\{j^{+}_{\mu5}(0)\, j_{\nu}^{em}(z) \right\}\! |\pi^{+}(p)\rangle\,;\la{eq:2a} \\
j^{+}_{\mu 5}(0) &=& \bar{d}(0) \gamma_{\mu} \gamma_5 u(0),~j_{\nu}^{em}(z)=e_u\bar{u}(z)\gamma_\nu u(z)+e_d\bar{d}(z)\gamma_\nu d(z) -\text{ the quark electromagnetic current}\,. \la{eq:2b}
\label{2:cor}
\ea
\end{subequations}
The correlator $T_{\mu\nu}$ serves to extract the emFF $F_\pi$  and can be presented as the expansions
both in twists and in PT series \cite{Bijnens:2002mg},
\begin{subequations}
\ba
T_{\mu\nu}(p,q)\! &\rightarrow& i f_\pi 2p_\mu p_\nu\left( T_\text{(tw2)}\otimes\varphi_{\pi}^\text{(2)} +
T_\text{(tw4)}\otimes\varphi_{\pi}^{(4)} + T_\text{(tw6)} \right)\,; \label{eq:3a}\\
 T_\text{(tw2)}(Q^2,\sigma; x)                  & =  &\sum_{k=0} a_s^k(\mu^2)\, T^\text{(k)}(x)\,, \label{eq:3b}
\ea
\end{subequations}
where the sign $\otimes$ means the convolution $A \otimes B = \int_0^1 dx A(x) B(x)$;\, $-q^2=Q^2, -(p-q)^2=\sigma$; the upper index $\text{(k)}$ numerates  orders of $a_s$.
Rearrangement  and summation by the renormalization group (RG) of the amplitude $T_\text{(tw2)}\otimes\varphi_{\pi}^\text{(2)}$ in Eq.(\ref{eq:3b}) are discussed in Sec.\ref{sec:theor-basis}, and some results for the LCSR amplitude $T^\text{(1)}(x)$ are outlined in Appendix \ref{App:A}.

Conceptually, the consideration here follows our previous analysis of a somewhat similar object -- the pion TFF in
\cite{Ayala:2018ifo,Ayala:2019etj,Mikhailov:2021znq,Mikhailov:2021cee},
with the goal to improve the results of  FOPT by involving a renormalization group and to extend the LCSR method to smaller transfer momentum $Q^2$.
In what follows we will hold the methodical correspondence between the pion TFF and the pion emFF.
Moreover, we will use nonperturbative inputs in our emFF calculations that correspond to the results for every twist in the phenomenological analysis for the pion TFF in \cite{Mikhailov:2021cee} because of their universal character.
It is instructive to follow  the general blockdiagram of this kind of calculations in Fig. \ref{fig:diagram},
wherein the standard FOPT LCSR approach is shown in the upper line for the comparison with ours in the lower line.
Below, we briefly discuss the contents of the blocks and their relations. \vspace*{-3mm}
 \vspace{-2mm}

The combined effect of RG summation and dispersion relation,
presented as blocks in the block diagram,
is the key transformation that lead to the new perturbation series within the framework of LCSR.
Specifically, the application of the dispersion relation to amplitudes that have already been RG-summed
(Sec.\ref{sec:theor-basis}) necessarily transforms the power series over the QCD running coupling $\bar{a}_s^\nu$,
which is defined as $\bar{a}_s=\bar{\alpha}_s/(4\pi)$, into a non-power series.
This transformation is discussed and derived in Section \ref{subsec:DR-tw-2}.
The new perturbation theory, which emerges from these actions,
reduces to the well-known fractional analytic perturbation theory (FAPT) with
three coupling constants $\{A_\nu, M_\nu, I_\nu\}$, which are the images of $\bar{a}_s^\nu$ \cite{Ayala:2018ifo,Ayala:2019etj,Mikhailov:2021znq,Mikhailov:2021cee}.
The FAPT goes back to the well-known paper \cite{Shirkov:1997wi} and
  has been further developed in \cite{Bakulev:2005gw,Bakulev:2006ex,Bakulev:2010gm}.
Charge $\A_{\nu}$ is the spacelike image of $\bar{a}_s^\nu$ \cite{Bakulev:2005gw}, $\M_{\nu}$ is the timelike one \cite{Bakulev:2006ex}, and $\I_{\nu}$ is their further generalization.
This kind of generalization of the coupling constant, $\I_\nu(y,x)$, arises due to the "subtraction of the continuum" action
\cite{Ayala:2018ifo,Mikhailov:2021znq}, which is characteristic of the standard LCSR method.
These FAPT coupling constants are bounded from above (except for a singularity near the origin at $\nu \leqslant 1$) and do not have Landau singularities at $\mu^2 = \Lambda^2_\text{QCD}$ \cite{Bakulev:2005gw,Bakulev:2006ex}, in contrast to the standard QCD coupling $\bar{a}_s(\mu^2)$.
For a detailed discussion and illustrations, see Appendix \ref{App:B}.
These features  of the FAPT couplings  allow one to extend the perturbative analysis well below  the conventional
pQCD boundary $\mu^2_0 = 1$ GeV$^2$.
At this point, we need to clarify the low-energy properties of theories based on dispersion relations, such as the FAPT,
in more detail.

As it follows from the second and third blocks  of our way,
see Fig.\ref{fig:diagram} (outlined blocks),
the new perturbation theory  appears almost automatically within LCSR
and can be expressed via the mentioned FAPT charges.
However, we also face some limitations of this approach.
To make it work at the index $\nu \leqslant1$, one must admit some additional constraints on the
FAPT charges at the origin $\mu = 0$.
These local constraints, which determine the charge behavior at the origin, are introduced and discussed in Sec.\ref{subsec:DR-tw-2},\,\ref{subsec:twist-4}, Appendix\ref{App:B}. They are not unique,
and other approaches to this problem modify the spectral density, as in \cite{Ayala:2017tco} and \cite{Ayala:2024ghk}.
We try to avoid introducing new specific scales (as well as explicit models) here.
To be more precise, in order to preserve and respect the holomorphic properties of physical quantities, we need to extend the lower limit of the spectral integration. In particular, if we change the QCD limit $-\Lambda^2_{QCD}$ to 0, keeping the QCD spectral density, we obtain the (F)APT coupling. If we fix the limit to $m_\pi^2$, we can parametrize this region through delta functions, as in \cite{Ayala:2012xf, Ayala:2017tco, Ayala:2024ghk}. If we fix the limit to $m^2_{gluon}-\Lambda^2_{QCD}$, keeping the QCD spectral density, we get the Massive Perturbation Theory (MPT) coupling \cite{Shirkov:2012ux}. We cannot guarantee the validity of FAPT at very low energies near the origin, but phenomenological studies, such as \cite{Khandramai:2011zd, Allendes:2014fua, Ayala:2018ulm}, show that it can be applied up to 0.5 GeV (without introducing an additional mass scale).
The various approaches to $\alpha_s$ at low energy are discussed in detail in the
comprehensive review \cite{Deur:2023dzc}.

The required important  technical results of FAPT are collected in Appendix \ref{App:B},
which can be considered as an independent compendium of the existing, useful FAPT results.
The formulas for the Borel transformation necessary to carry on the analysis are presented in Appendix \ref{App:C}.
The Borel images of a series in running $\bar{a}^n_s$ compared to $\mu$-ansatz  versus the $\mu$-ansatz \cite{Braun:1999uj,Bijnens:2002mg,Cheng:2020vwr}
(see  upper line in Fig.\ref{fig:diagram})
is discussed in part there.
The same approach  as for twist 2 is applied to the amplitude $T_\text{(tw4)} \otimes \varphi_{\pi}^\text{(4)}$,
of twist 4,  based on the additional necessary requirements discussed in Sec. \ref{subsec:DR-tw-2} and in Appendix \ref{App:D}.

In Sec. \ref{sec:data-fit}, we discuss important nonperturbative inputs for the LCSR (see also Appendix\,\ref{App:E}) that significantly affect the result.
Finally, we
provide our predictions for the emFF $F^{\text{LCSR}}_{\pi}(Q^2)$ in comparison with the experimental data.

\section{RG summation for the TFF and the {\scriptsize  em}FF  in the leading twist}
 \label{sec:theor-basis}
%
In order to carry out the RG summation, it is useful to expand the DA
$\varphi_{\pi}^{(2)}(x,\mu^2)$ of leading twist 2, as well as the corresponding
contribution to the em/transition amplitude over the
conformal basis of the Gegenbauer harmonics $\psi_n$, where
$\{\psi_n(x)=6x\bar{x}\,C^{3/2}_n(x-\bar{x}) \}$ and $b_n(\mu^2)$ are the coefficients of the conformal expansion,
\begin{subequations}
 \la{eq:gegen}
\ba
  \varphi_{\pi}^{(2)}(x,\mu^{2})&=& \psi_{0}(x)
  + \sum_{n=2,4, \ldots}^{\infty} b_{n}(\mu^{2}) \psi_{n}(x),
\la{eq:gegen-exp}\\
  T_\text{(tw2)}(Q^2,\sigma;x)\otimes\varphi_{\pi}^\text{(2)}(x)
&=&T_{0}(Q^2,\sigma)
\la{eq:gegen-FF} 
  + \sum_{n=2,4, \ldots}^{\infty} b_{n}(\mu^{2}) T_{(n)}(Q^2,\sigma;\mu^2)\,.
\ea
\end{subequations}
Here,  we follow the notations  \cite{Braun:1999uj,Bijnens:2002mg} for the emFF, taking into account Eq.(\ref{eq:2a}): $Q^2=-q^2$; $-(p-q)^2= -s=\sigma\geqslant 0$. The similar notations for the transition FF: $-q^2_1=Q^2$; $-q^2_2=q^2=\sigma\geqslant 0$ \cite{Ayala:2018ifo,Mikhailov:2021znq}; for all Bjorken fractions the bar means $\bar{x}=1-x, \bar{u}=1-u,\ldots$;
the lower index $n$ always numerates harmonics; $T_{(n)}$ is the projection of $T_\text{(tw2)}$ onto the basis vector $\psi_n$.
The description of both these form factors are relatively close one to another from the point of view of a general frame:
the initial correlators are  similar to one another and differ by the replacement of $\pi^{0}$ with $\pi^{+}$, the vector current  with the axial current at one of the verties,
they have practically the same diagrammatic presentations,  and their corresponding LCSRs are also close.
In particular, the  virtuality of the ``handbag'' diagrams, $ q(u)=\sigma u + Q^2 \bar{u}$, is the same for both amplitudes $T$,
and the corresponding evolutional logarithm $L(u)= \ln(q(u)/\mu^2)$ is the same.
For this reason, we will refer to the stages of calculation of the pion TFF that are already performed in \cite{Ayala:2018ifo,Mikhailov:2021znq}
 and discuss the similar stages of calculations for the emFF.

We transfer all the powers of the ERBL evolution logarithms, such as $L \cdot a_s V_0$,
$L \cdot a_s^2 V_1\ldots$, and $L \cdot a_s^2 \beta_0 \ldots$, which appear in higher orders of the amplitude $a^kT^{(k)}$ in (\ref{eq:3b}),
into a common exponential factor and the argument of $\overline{a}_s$,
as seen in Eqs. (\ref{eq:Fn-1loop-a}) and discussed in \cite{Ayala:2018ifo}.
This procedure goes \textit{beyond the FOPT} presented in the expansion (\ref{eq:3b}).
The remaining non-logarithmic terms $\mathcal{H}^\text{(1)}$ of partial amplitudes $T^\text{(k)}_{n}$ ($T^{(0)}= H^\text{(0)}$) are accompanied by the evolution  exponents for $\psi_{n}$--harmonics of DA expansion,
the corresponding exponentials were presented in the general form in
\cite{Ayala:2018ifo,Ayala:2019etj,Mikhailov:2021cee,Mikhailov:2021znq}.

Reducing the final result of the evolution to the one-loop level (\ie, taking only $V_0, \beta_0$),
we arrive at the compact expression on the right-hand side (rhs)
\begin{subequations}
 \label{eq:Fn-1loop}
\begin{eqnarray}
T_{(n)}(Q^2,\sigma;\mu^2)\stackrel{\text{1-loop}}{\longrightarrow} H_{(n)}^{(1)}
&=&
  H^\text{(0)}(u)
  \underset{u}{\otimes}
  \left[\1+ \bar{a}_s(u)\mathcal{H}^{(1)}(u,v)\right]
\exp\Bigl[\int_{a_s(\mu^2)}^{\bar{a}_s(u)} d\alpha
			\frac{\alpha V_0(v,x)}{\alpha^2\beta_0} \Bigr]
		~\underset{x}\otimes\psi_n(x)= \label{eq:Fn-1loop-a}\\
&=&H^\text{(0)}(u)
  \underset{u}{\otimes}
  \left[\1+ \bar{a}_s(u)\mathcal{H}^{(1)}(u,v)\right]
   \left(\frac{\bar{a}_s(u)}{a_s(\mu^2)} \right)^{\nu_n}
   \underset{v}\otimes
   \psi_n(v) \,  \label{eq:Fn-1loop-b} ,
\end{eqnarray}
 \end{subequations}
where the notation for the running coupling 
$\bar{a}_s(u)=\bar{a}_s( q(u))\equiv \bar{a}_s(\sigma u + Q^2 \bar{u})$ is used;
$H^\text{(0)}(u)$ is the Born term of the perturbative expansion, $\mathcal{H}^{(1)}(u,v)$
is the next-to-leading-order (NLO) coefficient function;
the ERBL kernel $V_0(v,x)$ is explicitly defined in Eq.~(\ref{eq:V})
and satisfies the equation,
\begin{eqnarray}
	V_0(v,x)\underset{x}{\otimes}\psi_n(x)&=&-\frac{1}{2}\gamma_0(n)\psi_n(u)\,,
~~\displaystyle\nu_n=\frac{1}{2}\frac{\gamma_0(n)}{\beta_0},
 \label{eq:various-c}
\end{eqnarray}
$a_s\gamma_{0}(n)$ -- the one-loop anomalous dimension of the
corresponding composite operator of the leading twist and $\beta_0$ is the first coefficient of the QCD $\beta$-function,
both defined in (\ref{eqA:RG-1l}).
The expression (\ref{eq:Fn-1loop}) coincides in structure with the corresponding RG-summed  expression for the TFF \cite{Ayala:2018ifo,Ayala:2019etj,Mikhailov:2021znq},
differing  from the latter by the forms of the amplitudes $H^\text{(0)}(u), \mathcal{H}^{(1)}(u,v)$
in comparison with the ones $T_0(u),\mathcal{T}^{(1)}(u,v)$ there.
The quantities entering into (\ref{eq:Fn-1loop}) are the following (for $\mathcal{H}^{(1)}$ we refer to Appendix \ref{App:A}):
\begin{subequations}
\label{eq:various}
\begin{eqnarray}
&&
	H^\text{(0)}(u) \equiv H^\text{(0)}(Q^2,\sigma;u)
	=
	\frac{u}{\sigma u + Q^2\bar{u}}\,;~
\\
&&
	\1 \equiv \delta(u-v)\,. 
\end{eqnarray}
\end{subequations}
Note, that  the running coupling
$\bar{a}_s^\nu(u) $
and the coefficient functions $H^\text{(0)}(u)$,
$\mathcal{H}^{(1)}(u,v)$ do not enter into (\ref{eq:Fn-1loop}) as a simple
product, but  as their convolution.
It is important to note that for the case of $\psi_0$ harmonic, the results of
RG summation in $H_{(0)}^{(1)}$ do not manifest themself due to the conservation of the vector (axial) current,
$\gamma_{0}(n=0)=0, \nu_{n=0}=0$ in Eqs.(\ref{eq:Fn-1loop}), (\ref{eq:various-c}), 
\be \label{eq:F0-1loop}
 H_{(0)}^{(1)}=
  H^\text{(0)}(u)
  \underset{u}{\otimes}
  \left[\1+ \bar{a}_s(u)\mathcal{H}^{(1)}(u,v)\right]
\underset{v}\otimes
   \psi_0(v)\,.
\ee
This circumstance may serve as a test for the current results.

For small values of $\sigma$, this convolution has
only a formal rather than a physical meaning, even at large $Q^2$.
This becomes obvious when the scale argument
$\sigma u+Q^2\bar{u}$ approaches small values for $u\to 1$, even if
$Q^2$ is large, so  the perturbative expansion  becomes unprotected.
 This problem is avoided when the amplitude $H_{(n)}^{(1)}$ is involved in
 the dispersion relation  in LCSR.
As we show below, in this case, an equation like Eq.\ (\ref{eq:Fn-1loop-b})
can still be safely used in the emFF calculation even for small $Q^2$.

\section{Dispersion relation in connection with the RG technique, the twist-2}
\label{sec:tw-2}
\subsection{Dispersion relation for the amplitude}
\label{subsec:DR-tw-2}
As we now demonstrate, summing over all radiative corrections in
Eq.\ (\ref{eq:Fn-1loop}) entails a new contribution to the imaginary part of the amplitude
$H_{(n)}(Q^2,-\sigma)$ and   generates a  new spectral density
$ \rho_n^\text{tw2}$, where $\sigma$ is dual to $s$.
This contribution marks an important difference from the standard version of the LCSRs
\cite{Khodjamirian:1997tk,Mikhailov:2009kf,Agaev:2010aq,Mikhailov:2016klg}.
To be more specific, the imaginary part of the Born contribution is induced by
the singularity of $ H_0(Q^2,-\sigma;u)$ multiplied by a power of the logarithms
that originated from the truncated pQCD series (usually a few terms) of radiative corrections.
In contrast, the RG summed radiative corrections lead to a term in the
spectral density $\rho_n^\text{tw2}$ that originates from the
$\Im_{s'}\left[\bar{a}^\nu_s(Q^2 \bar{u}- s'u)\right]/\pi$ contribution in the rhs of Eq.(\ref{eq:Fn-1loop}),
\begin{subequations}
 \label{eq:disp}
\be
 \rho_n^\text{tw2}(s',Q^2)= \Re\left[\frac{u}{\left(Q^2 \bar{u} - s'u\right)}\right]\frac{\Im_{s'}[\bar{a}^{\nu}_s(Q^2 \bar{u}- s'u)]}{\pi\, a_s^{\nu}(\mu^2_0)}\otimes \psi_n(u)+\bm{0}\,. \label{eq:tw2-rho}
 \ee
  The symbol $\bm{0}$ in the rhs of  (\ref{eq:tw2-rho}) means the trace of the previous ``standard'' \cite{Braun:1999uj,Bijnens:2002mg,Cheng:2020vwr} contribution that is now
 proportional to $\frac{1}{\pi}\Im\left( H_0\right)\,\Re[\bar{a}^{\nu}_s ]=0$ because
 the pole of the first multiplier at $Q^2 \bar{u}- s'u=0$
 coincides with the argument of the second one; at the same time, one should put $\bar{a}^{\nu}_s(0)=0$ formally.
 Let us introduce an auxiliary quantity $J_n$ to deal with the dispersion integral of expression  (\ref{eq:Fn-1loop}),
\be
J_n(Q^2,\sigma)= \int^{s_0}_0 \frac{ds'}{s'+\sigma}\, \rho_n^\text{tw2}(s',Q^2)\,. \label{eq:disp-tw-2}
\ee
 \end{subequations}
Note that the finite upper limit $s_0$ in the integral
(\ref{eq:disp-tw-2}) is the interval of duality \cite{Braun:1999uj,Bijnens:2002mg,Cheng:2020vwr},
while the integral by itself appears due to the ``subtraction of continuum'' inherent in the LCSR.
 Replacing the  integration variable $s',~s' \rightarrow s=s' u - Q^2 \bar{u}$ in (\ref{eq:disp-tw-2}) and taking into account that the integral exists only for $s\geqslant 0$, one obtains
\begin{subequations}
 \ba \label{eq:F2-SR}
 J_n(Q^2,\sigma)&=& \frac{1}{a_s^{\nu_n}(\mu^2_0)}\left(\int^{s(u)}_0\frac{ds}{s}\frac{-u}{(s+q(\bar{u}))}\cdot\frac{1}{\pi}\Im[\bar{a}^{\nu_n}_s(-s-i\varepsilon)]\, \right)
 \otimes \left(\, \theta(u \geqslant u_0) \psi_n(u)\right), \\
\text{where the upper limit}&&\, s(u)= (Q^2 +s_0)( u-u_0)\geqslant 0,\, u_0=Q^2/(Q^2+s_0);~ q(u)= \sigma \,u + Q^2\, \bar{u}\,.
 \ea
   \end{subequations}
 In the rhs of (\ref{eq:F2-SR}), there appears the spectral density $\rho_\nu$ for $\bar{a}_s^{\nu}$,
\begin{subequations}
\be\label{eq:key_disper}
  \Ds \rho_\nu(s) =\frac{1}{\pi}\Im[\bar{a}^{\nu}_s(-s-i\varepsilon)]\,,
\ee
 which is the key quantity of FAPT, see also its expression in Eq.(\ref{eq:spectr-dens}) in Appendix \ref{App:B}.
After a simple algebra with the dispersion integral in the parentheses in (\ref{eq:F2-SR}) and taking into account the
definitions of FAPT couplings $\A_{\nu}, \I_{\nu}$ in Eq.(\ref{eq:faptcouplings}) and (\ref{eq:A-def}) respectively, 
one arrives at the expression
\ba\label{eq:1subtractions}
&&\frac{1}{x} \left(\int_0^y=\int_0^\infty ds - \int_y^\infty ds\right) \left(\frac{1}{s+x} -\frac{1}{s}\right)\rho_\nu(s) =
\frac{1}{x}\left[\left(\A_{\nu}(x)-\A_{\nu}(0)\right) - \left(\I_{\nu}(y,x)-\I_{\nu}(y,0)\right)\, \right]\,.
\ea
 \end{subequations}
Hereinafter we use the notation $\Ds x=q(u)$ and $y= s(u)$ for brevity.
Substituting (\ref{eq:1subtractions}) in (\ref{eq:F2-SR}), we arrive at the final expression for $J_n$,
\ba \label{eq:H2}
&&\!\!\!\!J_n(Q^2,\sigma)\!=\!\!\frac{u}{x}\frac{1}{a_s^{\nu_n}(\mu^2_0)}\!
\left[\A_{\nu_n}(x)-\A_{\nu_n}(0)- \left(\I_{\nu_n}(y,x)-\I_{\nu_n}(y,0)\right)
\right]\bigg|_{\begin{subarray}{c}x=q(u)\\y= s(u)\end{subarray}}\!\!\otimes \!\left(\theta(u \geqslant u_0) \psi_n(u)\right),
\ea
where we find the result of the dispersion relation ``with one subtraction''.
Here, we introduce an effective coupling constant $\Delta_{\nu}$ that appears in square brackets in Eq. (\ref{eq:H2}),
 and can be rewritten as
\be \label{eq:effective}
 \Delta_{\nu}(y,x)= \I_{\nu}(0,x)- \I_{\nu}(y,x)+\I_{\nu}(y,0)-\A_{\nu}(0) =\A_{\nu}(x)- \I_{\nu}(y,x) + \M_{\nu}(y)-\A_{\nu}(0),
\ee
The generalized charge $\I_{\nu}(y,x)$, which reduces to $\I_{\nu}(0,x)=\A_{\nu}(x)$, $\I_{\nu}(y,0)=M_{\nu}(y)$, is defined in Eq. (\ref{eq:B8}).
Following to Eq.(\ref{eq:H2},\ref{eq:effective}) the  particular $H_{(n)}^{(2)}$ in (\ref{eq:Fn-1loop}) is transformed using the dispersion
approach to
\ba
 \label{eq:F2dispers}
H_{(n)}^\text{(1)LCSR}=\frac{1}{a_s^{\nu_n}(\mu^2_0)}H_0(u)\bigg\{ \Delta_{\nu_n}\left(s(u),q(u)\right)\underset{u}\otimes \1
 &+& \Delta_{(1+\nu_n)}\left(s(u),q(u)\right)\underset{u}\otimes \mathcal{H}^{(1)}(u,v)\bigg\}\underset{v}\otimes\!(\theta(v \geqslant v_0)\psi_n(v))\,,
\ea
In Eq.(\ref{eq:F2dispers}), the effective coupling constant $\Delta_\nu$ replaces the RG-summed QCD coupling constant $\overline{a}_s^\nu$ in Eq. (\ref{eq:Fn-1loop}), Sec.\ref{sec:theor-basis}.

Until now, our derivations were based on explicit QCD calculations within the LCSR framework.
This led us to the FAPT approach, which has a singular point at the origin,
where the derivatives diverge for indices $0 < \nu \leqslant 1$.
Let us mention that $A_\nu(0) = M_\nu(0)=0$ hold  in itself for the domain $\nu > 1$,
as seen in Eqs.(\ref{eq:couplings}).
We found that \textit{additional conditions on $A_\nu(0)=M_\nu(0)=0$,
at any $\nu$}, are necessary to ensure the consistency of the LCSR results for TFF
\cite{Ayala:2018ifo,Mikhailov:2021cee}.
These conditions will be consequently applied here to the results for emFF, Eq.(\ref{eq:A-constraint}),
therefore for $\Delta_{\nu}$ we obtain
\be \Delta_{\nu}(x,y)=\A_{\nu}(x)- \I_{\nu}(y,x) + \M_{\nu}(y). \ee
The cumbersome three-term expression for $\Delta_\nu$ in Eq. (\ref{eq:effective})
is due to the ``subtraction of the continuum'' that is inherent in LCSRs. Indeed,
when $s_0$ approaches infinity, the effective charge $\Delta_\nu$ tends to $\A_\nu(x)$,
that is, it tends to the spacelike FAPT coupling constant, as expected.

So, the final result of the LCSR (twist 2) can be obtained  by
employing a linear combination of generalized FAPT couplings $\I_{\nu}$ collected
into a single effective $\Delta_{\nu}$.
For the important partial case of asymptotic DA $\psi_0$, at $n=0, \nu_{n=0}=0$,  this effective coupling
$\Delta_{\nu}$ turns  into
\begin{subequations}
\ba \label{eq:delta}
 \Delta_{0}(y,x)= \A_{0}(x)- \I_{0}(y,x)+\M_{0}(y) \rightarrow 1;~\Delta_{1}(y,x)= \A_{1}(x)- \I_{1}(y,x)+\M_{1}(y)
 \,,
  \ea
at that $\A_{1}(x), \M_{1}(y)$ are expressed in terms of elementary functions \cite{Bakulev:2006ex}, see Eq.(\ref{eq:couplings}).
Finally, for the contribution of $\psi_0$-harmonics to $H^\text{(2)LCSR}$ we obtain
\be
 \label{eq:H0dispers}
H_{(0)}^\text{(1)LCSR}=H^\text{(0)}(u)\underset{u}\otimes \bigg\{ \1
 + \Delta_{1}(s(u),q(u)) \mathcal{H}^{(1)}(u,v)\bigg\}\underset{v}\otimes\!(\theta(v \geqslant v_0)\psi_0(v))\,.
\ee
 \end{subequations}
It should be compared with Eq.(\ref{eq:F0-1loop}) that was obtained at the same condition before the dispersion relation is used:
the single change is the replacement of the standard $\bar{a}_s(u)$ with the new coupling $\Delta_{1}(s(u),q(u))$ in Eq.(\ref{eq:H0dispers}).
Concluding, one can present the chain of all the previous transformations with coupling constants as

\centerline{\begin{tabular}{lll|l}
FOPT                      & RG resum, Eq.(\ref{eq:Fn-1loop}):                                   &LCSR for $F_{\pi}$, Eq.(\ref{eq:effective}):               &~ \\ \hline
                          &                                            & dispersion relat. +       &Only dispersion relat. \\               &                                            &duality model with $s_0$   &at $~s_0 \rightarrow\infty$ \\                          $a_s^n(\mu^2)\longrightarrow$&~$\bar{a}_s^\nu(x=\sigma u+Q^2\bar{u})\longrightarrow$&\hspace{7mm}$\Delta_{\nu}(y,x)~\Longrightarrow$&~$\A_\nu(x=\sigma u+Q^2\bar{u})$\,. \\ \hline
\end{tabular}
}
\subsection{ Contribution of the Borel image to the LCSR}
 \label{sec:borel}
The further application of the Borel transform $M^2\mathbf{\hat{B}}_{(\sigma\to M^2)}$ to the amplitude $H^{(1)\text{LCSR}}_n$ in order to obtain the completed contribution $F^{(1)\text{LCSR}}$ in the LCSR violates the universality of the coupling constant $\I_\nu(y,x)$, as manifested in Eqs.(\ref{eq:F2dispers}) and (\ref{eq:H0dispers}). This results in various forms of the FAPT coupling constants after the Borel transform.
So, applying $ M^2 \mathbf{\hat{B}_{(\sigma \to M^2)}}$ to $\Ds H^\text{(0)}(u)\Delta_{\nu}(s(u),q(u))$ and using (\ref{eq:B4b}), (\ref{eq:key_disper}) and (\ref{eq:B7}), one arrives at
\begin{subequations}
 \label{eq:F2LCSR}
\ba
\!\!\!\!\!\!\!\!F^\text{(1)LCSR}_{\pi,n}(Q^2,M^2)&\!=\!&\!\! M^2 \mathbf{\hat{B}}H_{(n)}^\text{(1)LCSR}
 \! =\!\!\frac{1}{a_s^{\nu_n}(\mu^2_0)}\Bigg\{ \left[\M_{\nu_n}\left(s(u)\right)\!+\int^{s(u)}_0\!\!\!\!\rho_{\nu_n}(s) \omega_1(s,u)ds\right]\!\underset{u}\otimes \1+ \label{eq:F2LCSR-a}\\
  &&\!\!\!\! \left[\M_{(\nu_n+1)}(s(u))+\!\int^{s(u)}_0\!\!\!\!\rho_{(\nu_n+1)}(s) \omega_1(s,u)ds\right]\!\underset{u}\otimes \mathcal{H}^{(1)}(u,v)\Bigg\}
  \exp{\Ds \left(\!-\frac{Q^2}{M^2}\frac{\bar{u}}{u}\right)}\!
\underset{v}\otimes \!(\theta(v \geqslant v_0)\psi_n(v)),  \label{eq:F2LCSR-b}
\ea
\end{subequations}
where the weight $\Ds \omega_1(s,u)=\frac{1}{s}\left[1-\exp{\left(-\frac{s}{M^2 u}\right)} \right]$, $v_0=u_0$.

Equation (\ref{eq:F2LCSR-a}) provides the contribution of twist-2 for pion's emFF in the framework of the LCSR at RG summation at LO.
Taking into account that $\I_\nu(s(u),0)=\M_\nu(s(u)))\Big|_{\nu \to 0}=1$ for the first term, and $\rho_{(\nu)}\Big|_{\nu \to 0} \to 0$ for the second one in square brackets, one arrives
at the representation of Eq.(\ref{eq:F2LCSR-a}) at different $n$,
\begin{subequations}
\label{eq:LO-LCSR}
\ba
\!\!\!\!\!\!\!F^\text{(1)LCSR,LO}_{\pi,0}(Q^2,M^2)&=&
\Ds \int^1_{u_0} \exp{\Ds \left(\!-\frac{Q^2}{M^2}\frac{\bar{u}}{u}\right)} \psi_0(u) du,\,(n=0); \label{eq:LO-LCSR-a}\\
\!\!\!\!\!\!\!F^\text{(1)LCSR,LO}_{\pi,n>0}(Q^2,M^2)&=& \frac{1}{a_s^{\nu_n}(\mu^2_0)}\!\! \left[\M_{\nu_n}\!\left(s(u)\right)\!+ \! \Ds \int^{s(u)}_0\!\!\!\!\rho_{\nu_n}(s) \omega_1(s,u)\,ds\right]\exp{\Ds \left(\!-\frac{Q^2}{M^2}\frac{\bar{u}}{u}\right)}\!\underset{u}\otimes \!\left(\theta(u \geqslant u_0)\psi_n(u)\right),\,
 \label{eq:LO-LCSR-b}
\ea
\end{subequations}
where the $\psi_0$-projection in the first line, Eq.(\ref{eq:LO-LCSR-a}), coincides with the standard FOPT \cite{Braun:1999uj,Bijnens:2002mg,Cheng:2020vwr}
result due to the conservation of the vector current.
The expressions for higher harmonics contain the results of RG-summation within the square brackets in the second line, Eq.(\ref{eq:LO-LCSR-b}), while the complete  Eq.(\ref{eq:F2LCSR}) provides the NLO contribution in addition to Eq.(\ref{eq:LO-LCSR-b}).

\section{Contributions of higher twists 4,\,6 to the LCSR}
 \label{sec:higher_twist}
 \subsection{Twist-4 contribution via the dispersion relation}
   \label{subsec:twist-4}
As was mentioned above, the twist-4 contribution is crucial for improving precision at moderate $Q^2$.
 This contribution becomes comparable with the twist-2 one
  at small $Q^2 \sim 0.5$\,GeV$^2$ 
  see the discussion in \cite{Bijnens:2002mg}.
 The  amplitude corresponding to this contribution reads \cite{Bijnens:2002mg}
\begin{subequations}
 \begin{eqnarray} \label{eq:tw4-amplitude}
 T_\text{(tw4)}(Q^2,-s)=\left(\frac{1}{\bar{u}\,Q^2-u\,s}\right)^2 \delta_\text{tw-4}^2(\mu^2)\,\otimes (u\varphi^{(4)}_\pi(u))\,;
 \end{eqnarray}
 \begin{eqnarray}
	&&
	\delta^2_\text{tw-4}(\mu^2)
	= \left[\frac{a_s(\mu^2)}{a_s(\mu^2_0)}
	\right]^{\nu_{t4}}\delta_\text{tw-4}^2(\mu^2_0)\,,
	\nu_{t4}=\gamma_{t4}/\beta_0\, ,\gamma_{t4} = C_\text{F}\,8/3\, ,\nu_{t4}=C_\text{F} 8/(3\beta_0) < 1; \label{eq:tw4-RG-b}\\
&&
	\delta^2_\text{tw-4}(\mu^2_0) = 0.21~\text{GeV}^2\,,  \varphi^{(4)}_\pi(u)= \frac{20}3 u\bar{u}\left(1-u(7-8u)\right)\,.\label{eq:tw4-RG-c}
\end{eqnarray}
\end{subequations}
The shape of DA $\varphi^{(4)}_\pi$ in (\ref{eq:tw4-RG-c}) means the asymptotic form of all twist 4 DAs, which form it.
The spectral density $\rho^\text{tw4}$ taken on the variable $s$ for the amplitude (\ref{eq:tw4-amplitude}) is determined by the imaginary part that appears only due to the first factor there.
 Thus $\rho^\text{tw4}(s)$ is proportional to $\Ds \left[\delta(\bar{u}/u\,Q^2-s)\right]'_{s}$, which leads to the known  expression \cite{Bijnens:2002mg,Cheng:2020vwr} that is also derived in Appendix\,\ref{App:D}.

Here we suppose an alternative approach, based on the assumption that two-particles DAs of the twist 4, corresponding to the ``hand bag'' diagram provide the main contribution to the final  $\varphi^{(4)}$. Next, we suppose  that the virtuality  $\mu^2$, related to this diagram, can be taken as  $\bar{u}\,Q^2-u\,s = \mu^2$, its typical scale.  This hypothesis leads to $a_s(\mu^2) \to a_s(\bar{u}\,Q^2-u\,s )$, which in turn, leads to the involvement of latter into the dispersion relation.  Therefore, the use of such   hypothesis completely changes  spectral density $\rho^\text{tw4}(s')$ and makes the situation similar to the consideration in Sec. \ref{sec:tw-2}.
In other words, \textit{the assumption} of RG running taken together with the dispersion relation for the twist 4 leads to a non-power series and the appearance of coupling constants like  FAPT in full analogy with the results in the leading twist.
Following the procedure used in Sec.\,\ref{sec:tw-2}
(elaborated in \cite{Ayala:2018ifo,Ayala:2019etj,Mikhailov:2021znq,Mikhailov:2021cee} ), see Eq.(\ref{eq:disp}) there, one obtains
\begin{subequations}
 \begin{eqnarray}
 \rho^\text{tw4}(s',Q^2)&=& \Re\left[\frac{1}{\left(Q^2 \bar{u} - s'u\right)^2}\right]\frac{1}{\pi }\Im[\bar{a}^{\nu_{t4}}_s(Q^2 \bar{u}- s'u)]\otimes \left(u \varphi^{(4)}_\pi(u)\right)+\bm{0}\,, \label{eq:tw4-rho} \\
 H^\text{LCSR}_\text{(tw4)}(Q^2,\sigma)&=&\frac{\delta^2_\text{tw-4}(\mu^2_0)}{a_s^{\nu_{t4}}(\mu^2_0)}\, \int^{s_0}_0 \frac{ds'}{s'+\sigma}\, \rho^\text{tw4}(s',Q^2), \label{eq:disp-tw-4}
 \end{eqnarray}
  \end{subequations}
 where $\sigma =-s\geqslant 0$.
 The symbol $\bm{0}$ in the rhs of  (\ref{eq:tw4-rho}) means the trace of the previous FOPT contribution that is now
 proportional to $\frac{1}{\pi}\Im\left( H_0^2\right)\,\Re[\bar{a}^{\nu}_s ]=0$, like for the case of twist 2.
 Replacing the  integration variable $s',~s' \rightarrow s=s' u - Q^2 \bar{u}$ in (\ref{eq:disp-tw-4}) and taking
 into account that the dispersion integral exists only for $s\geqslant 0$, one obtains
 \ba \label{eq:F4-19}
 H^\text{LCSR}_\text{(tw4)}(Q^2,\sigma)\!&\!=\!&\!\frac{\delta_\text{tw-4}^2(\mu^2_0)}{a_s^{\nu}(\mu^2_0)}\, \left(\int^{s(u)}_0\frac{ds}{s^2}\frac{1}{(s+q(\bar{u}))}\,\frac{1}{\pi}\Im[\bar{a}^{\nu}_s(-s-i\varepsilon)]\right)
 \otimes \left(u\, \theta(u \geqslant u_0) \varphi^{(4)}_\pi(u)\right),
 \ea
where the upper limit $s(u)= (s_0+Q^2)(u-u_0);\,s_0\geqslant s(u)\geqslant 0;\, u_0=Q^2/(Q^2+s_0);~ q(u)= \sigma \,u + Q^2\, \bar{u}$\,.
Performing a simple algebra with the dispersion integral in the first parentheses in (\ref{eq:F4-19}) and using (\ref{eq:key_disper}), one arrives at the expression
\ba\label{eq:2subtractions}
&&\frac{1}{x^2} \left(\int_0^\infty ds - \int_y^\infty ds\right) \left(\frac{1}{s+x} -\frac{1}{s}+\frac{x}{s^2}\right)\rho_\nu(s) =
 \nonumber \\
&&\frac{1}{x^2}\left[\A_{\nu}(x)-\A_{\nu}(0)-x \A'_{\nu}(0) \right]- \frac{1}{x^2}\left[\I_{\nu}(y,x)-\I_{\nu}(y,0)-x \I'_{\nu}(y,0)\, \right]\,,
\ea
where $\Ds \I'_{\nu}(y,0)\equiv \frac{d}{dx}\I_{\nu}(y,x)\Big{|}_{x=0}$ means the differentiation by the second argument
(recall that $x=q(u)$, $y= s(u)$).
In Eq.(\ref{eq:2subtractions}) for twist 4 we get the dispersion relation with two subtractions in contrast to twist 2 result with a single subtruction in (\ref{eq:H2}).\!\!\!\!
Again, taking into account Eq.(\ref{eq:2subtractions}), we introduce an effective charge $\Box_{\nu}$ that is typical for twist 4,
\be
 \label{eq:Box}
\Box_{\nu}(y,x) = \A_{\nu}(x)-\I_{\nu}(y,x)-\A_{\nu}(0)-x \A'_{\nu}(0)+\M_{\nu}(y)+x \I'_{\nu}(y,0) \,.
\ee
Substituting Eqs.(\ref{eq:2subtractions}),(\ref{eq:Box}) in Eq.(\ref{eq:F4-19}), one obtains the final result for $H^\text{LCSR}_\text{(tw4)}$,
\ba
H^\text{LCSR}_\text{(tw4)}(Q^2,\sigma) &=& \frac{\delta_\text{tw-4}^2(\mu^2_0)}{(\sigma u+Q^2\bar{u})^2}\frac{1}{a_s^{\nu_{t4}}(\mu^2_0)}
\Box_{\nu_{t4}}(s(u),q(u))\!\otimes \!\left(u\theta(u \geqslant u_0) \varphi^{(4)}_\pi(u)\right).\label{eq:H4FAPT}
\ea
In Eq.(\ref{eq:Box}), we encounter the ill-defined FAPT charges $\A_{\nu}(0), \A'_\nu(0)$ at the origin:
these charges diverge at $\nu_{t4}<1$ (see (\ref{eq:tw4-RG-b})).
Taking into account the necessary condition \cite{Ayala:2018ifo} $\A_\nu(0) = \M_\nu(0)=0$
we need to extend these constraints to the first derivatives:
$\A'_\nu(0)=\M'_\nu(0)=0$, as we previously applied in \cite{Mikhailov:2021cee}.
We required this to make the TFF calculation self-consistent.
In general, to solve the ill-defined problem, we may need to use a sufficiently high-order zero
for the FAPT charge at the origin.
Considering this condition and applying the Borel transform to $H_\text{LCSR}^{(\text{tw4})}$ in (\ref{eq:H4FAPT}),
we obtain the contribution from $F_\pi^{(\text{tw}4)}$,
\begin{subequations}
\ba \label{eq:F4a}
\!\!\!\!\!\!F^\text{(tw4)}_\pi(Q^2,M^2)&=& M^2 \mathbf{\hat{B}_{(\sigma \to M^2)}}\left\{
\frac{\delta_\text{tw-4}^2(\mu^2_0)}{(\sigma u+Q^2\bar{u})^2} \frac{1}{ a_s^{\nu_{t4}}(\mu^2_0)}\Box_{\nu_{t4}}(s(u),q(u))
\!\!\otimes \!\!\left(u \theta(u \geqslant u_0) \varphi^{(4)}_\pi(u)\right)\!\right\}, \\
\text{where}~ \Box_{\nu}(y,x) &=& \A_{\nu}(x)-\I_{\nu}(y,x)+\M_{\nu}(y)+x \I'_{\nu}(y,0)\,.
\ea
 \end{subequations}
We conclude that the amplitude $H^{\text{LCSR}}_{(\text{tw}4)}$ can be expressed
in terms of an effective coupling $\Box_\nu$ that consists of generalized FAPT
couplings $I_\nu$ and their derivatives, in full analogy with the previous case of TFF \cite{Ayala:2018ifo,Mikhailov:2021znq,Mikhailov:2021cee}.
Moreover, if the single parameter $s_0$ of the LCSR tends to infinity, then the effective
coupling $\Box_\nu$ tends towards the spacelike $\A_{\nu}(x)$, as expected.

The subsequent  Borel transform changes this correspondence, creating new constructions.
Indeed, applying Eq.(\ref{eq:B3}) to those terms that related to last two in (\ref{eq:Box}), and  Eq.(\ref{eq:B4})
to the first difference $\A_{\nu}(x)\!-\! \I_{\nu}(y,x)$ there, we arrive at
\ba \label{eq:F4b}
\!\!\!\!F^\text{(tw4)LCSR}_\pi(Q^2,M^2)=&&\frac{\delta_\text{tw-4}^2(\mu^2_0)}{a_s^{\nu_{t4}}(\mu^2_0)}\cdot \\
&&\!\!\Ds \left[\!\M_{\nu_{t4}}(s(u))\!+\!\frac{1}{M^2 u} \I'_{\nu_{t4}}(s(u),0)\!+\!\int^{s(u)}_0\!\!\!\!\!\!\!\rho_{\nu_{t4}}(s) \omega_2(s,u)ds\right]\! \exp{\Ds \left(\!-\frac{Q^2}{M^2}\frac{\bar{u}}{u}\right)}\!
\otimes \!(\theta(u \geqslant u_0)\varphi^{(4)}_\pi(u))\,, \nonumber
\ea
where the weight $\Ds \omega_2(s,u)=\frac{1}{s^2}\left[\exp{\left(-\frac{s}{M^2 u}\right)}-\left(1-\frac{s}{M^2 u}\right) \right]$ contains two subtractions of the  expansion of the exponential.
The integration of the second term $\sim \I'_{\nu}$ in (\ref{eq:F4b}) can be performed with the help of Eqs.(\ref{eq:B9}, \ref{eq:key-int},
\ref{eq:final-key-int}) in Appendix\,\ref{App:B}.

\begin{figure}[h]
\includegraphics[width=0.5\textwidth]{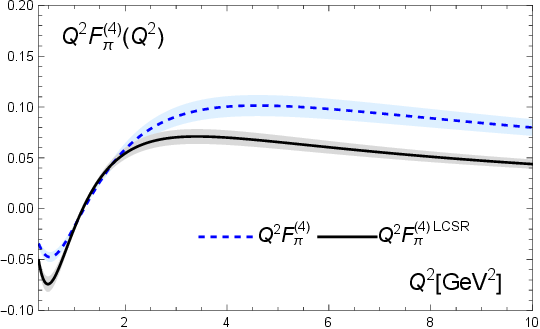}
 \caption{\footnotesize{\label{fig:tw-4}
Solid (black) line is the result of RG sum $F^\text{(4)LCSR}_\pi$ in Eq.(\ref{eq:F4b}); the dashed (blue) line is the standard result $F^{(4)}_\pi$ in Eq.(\ref{eq:standF4}) with the value of $\delta_\text{tw-4}^2(\mu^2_0)$ in \cite{Bijnens:2002mg}, both curves are taken at $M^2=1.2$\,GeV$^2$. }}
\end{figure}

The RG (summed)  result in (\ref{eq:F4b}) should be compared with the initial unsummed expression for $F^{(tw4)}_\pi(Q^2,M^2)$ (or one that is presented in (\ref{eq:finalF4})),
\ba \label{eq:standF4}
F^{(4)}_\pi(Q^2,M^2) 
&=&\delta_\text{tw-4}^2(\mu^2)\cdot \left[\delta(u -u_0) \frac{u}{Q^2} +\frac{1}{M^2 u} \right]
                         \exp{\Ds \left(- \frac{Q^2}{M^2}\frac{\bar{u}}{u}\right)} \otimes (\theta(u\geqslant u_0) \varphi^{(4)}_\pi(u))\,.
\ea
Here the authors of \cite{Braun:1999uj,Bijnens:2002mg,Cheng:2020vwr} used an ad hoc ansatz for the normalization
scale $\mu^2 \to \mu^2_u = M^2\,u + Q^2\,\bar{u}$;
see the discussion of it in Appendix\,\ref{App:C}.
The comparison of both  results for twist-4  are presented in Fig.\ref{fig:tw-4},
where for $\delta_\text{tw-4}^2(\mu^2_0)$ in Eq.(\ref{eq:F4b}),
we used the numerical estimate based on the relation of the latter with the parameter $\lambda^2_q/2$ of nonlocal condensate (NLC) \cite{Bakulev:2002uc}; see the beginning of Sec.\ref{sec:data-fit1}.
This parameter simultaneously determines the features of the twist-2 pion DA in the framework of the NLC QCD SR \cite{Bakulev:1995ck,Bakulev:2001pa}.
So, the twist-2 and twist-4 contributions of are mutually related  by means of the influence on $\varphi_{\pi}^{(2)}(x,\mu^2)$,
its  Gegenbauer expansion coefficients $\{b_n(\mu^2_0) \}$ in Eq.(\ref{eq:gegen}),
and  the coefficient $\delta_\text{tw-4}^2(\mu^2_0)$.
Note the important role of condensate nonlocality was shown in the calculation of emFF within QCD sum rules  in \cite{Bakulev:2009ib}.

As is seen in Fig.\,\ref{fig:tw-4},  the result of $F^\text{(tw4)LCSR}_\pi$ in  Eq.(\ref{eq:F4b})
is almost everywhere expectedly  lower than the standard $F^{(4)}_\pi$ in Eq.(\ref{eq:standF4}).
The behavior with $Q^2$ for both cases is close to each other, although the origins of the corresponding spectral densities are different.
Moreover, $F^\text{(tw4)LCSR}_\pi$ is well founded at small $Q^2 < 1$\, GeV$^2$ from the view point of radiative corrections.
Nevertheless, one should not fall below the scale $Q^2_{thr}\approx 0.4 - 0.5$\,GeV$^2$ -- the minimum of $Q^2 F^\text{(tw4)LCSR}(Q^2)$ for another reason; starting with this scale, the twist hierarchy becomes violated.

\subsection{Twist-6 contribution}
 \label{subsec:twist-6}
The factorizable twist-6 contribution \cite{Braun:1999uj} is
\ba
F^\text{(tw6)}_\pi(Q^2)=\frac{4\pi C_\text{F}}{3f_\pi^2\,Q^4}~\delta_\text{tw-6}^2(\mu^2_0)\,,&& \text{where} \label{eq:tw-6} \\
\delta_\text{tw-6}^2(\mu^2_0=1\,\text{GeV}^2)&=&
   \langle\sqrt{\alpha_s}\bar{q} q\rangle^2=\bm{(1.61\pm 2\cdot 0.26)}\,\times 10^{-4}\,\text{GeV}^6 \text{-- ``the best fit''
       in \cite{Mikhailov:2021znq}}\label{eq:val-tw-6}\\
 &&\langle\sqrt{\alpha_s}\bar{q} q\rangle^2=\left(~1.84^{+0.84}_{-0.24}~\right)\,\,\times 10^{-4}\,\text{GeV}^6\, \text{\cite{Cheng:2020vwr}}\,, \nn
\ea
Further, we will use our best-fit estimate
\footnote{ we recognize now that the conservative uncertainty should be approximately twice as large as that mentioned in \cite{Mikhailov:2021znq}}
 in Eq.(\ref{eq:val-tw-6}), which overlaps within errors with values given in \cite{Cheng:2020vwr}.
\section{Numerical estimates of $F_\pi$ in different LCSR approaches}
 \label{sec:data-fit}
 Firstly, we discuss in detail the status of nonperturbative inputs of different twists that significantly affect the resulting estimates of $F_\pi$ within the LCSRs.
 Then, in the next subsections, we will consider our predictions for $F_\pi$ at different approximations in comparison with experimental data.
 \subsection{Nonperturbative inputs of the LCSR} 
  \label{sec:data-fit1}
 In items (\textbf{i - iv}) below, we discuss 
 the admissible  models of the  DA of twist-2 and the factors at twists-4 and -6 that are compatible with the results of lattice calculation \cite{Bali:2021qem} and the phenomenological analysis of experimental data on TFF in \cite{Mikhailov:2021znq}.
Further, for duality interval $s_0$ of the LCSR we admit, the estimate $s_0=0.7$\,GeV$^2$\, \cite{Braun:1999uj,Bijnens:2002mg,Cheng:2020vwr}.

\textbf{(i)} We will use the corrected domain of the Gegenbauer expansion coefficients for the  pion DA $\varphi_{\pi}^{(2)}(x,\mu^2_0=1\text{GeV}^2 )$
 -- $\left(b_2^\text{BMS}(\mu^2_0)=\bm{0.159}^{+0.025}_{-0.027}, b_4^\text{BMS}(\mu^2_0)=\bm{-0.098}^{+0.05}_{-0.03}\right)$.
 Such modified BMS (BMSmod) model $\varphi^{(2)\text{BMS}}_\pi$ represents a compromise of QCD SR \cite{Bakulev:2001pa,Bakulev:2004mc} results,
 lattice simulation results \cite{Bali:2021qem}, and phenomenological analysis of the pion TFF\,\cite{Mikhailov:2021znq},   see Fig.1 there,
 which are reproduced as Fig.\,\ref{fig:fitTFF-NNLO} in Appendix \ref{App:E}.
 The vertices of the rectangle that bounds the uncertainty domain $(b_2,b_4)$  are:
  $$\{(b_2=0.132, b_4=-0.06), (b_2=0.185, b_4=-0.08), (b_2=0.185, b_4=-0.16), (b_2=0.132, b_4=-0.12)\}.$$
 The different lattice estimates \cite{Bali:2021qem} read $b_2(\mu^2_{L})=0.116\pm0.020$ (N$^3$LO) at $\mu^2_{L}=(2$ GeV$)^2$,
 which just coincide with the BMSmod component $b_2^\text{BMS}(\mu^2_0)=\bm{0.159}$ in \cite{Mikhailov:2021znq}
and  agrees with other lattice estimates $(b_2(\mu^2_{L})=0.131\pm 0.041,~b_4(\mu^2_{L})=-0.39\pm 0.76)$ presented in \cite{Detmold:2023voi}.

\textbf{(ii)} The estimate of the twist-4 parameter $\delta_\text{tw-4}^2(\mu_0^2=1\,\text{GeV}^2) \approx \lambda^2_q/(2\cdot 1.075)$ which is used here, is taken from QCD SR in \cite{Bakulev:2002uc}.
 The estimate of the ``nonlocality parameter'' $\lambda^2_q$ \cite{Mikhailov:1988nz,Mikhailov:1991pt} is $\lambda^2_q \in [0.4 - 0.45]$\, GeV$^2$ \cite{Mikhailov:2021znq}, that leads to $\delta_\text{tw-4}^2=\bm{0.198\pm 0.01\pm 0.01}$ GeV$^2$
(where the first uncertainty is due to variation of $\lambda^2_q$  and the second one to  the analysis of QCD SR \cite{Bakulev:2002uc} for $\delta_\text{tw-4}^2$),
 while the previous estimates were $\delta_\text{tw-4}^2(\mu_0^2)=0.17, 0.18$ GeV$^2$ that correspond to \cite{Bijnens:2002mg}, \cite{Cheng:2020vwr} respectively.

\textbf{(iii)} An alternative to the BMSmod DA is the so-called ``platykurtic'' DA \cite{Stefanis:2015qha}
$(b_2^{pk}=0.081, b_4^{pk}=-0.019)$ that corresponds to
the highest estimate of  $\lambda^2_q=0.45$\,GeV$^2$, see Fig.\ref{fig:fitTFF-NNLO} that leads to $\delta_\text{tw-4}^2=0.209\pm 0.01\pm 0.01$ GeV$^2$ \cite{Mikhailov:2021znq}.

\textbf{(iv)} The value of the scale parameter for tw-6 in Eq.(\ref{eq:tw-6}) is taken from the results \cite{Mikhailov:2021znq}
$ \langle\sqrt{\alpha_s}\bar{q} q\rangle^2=\bm{(1.61\pm 2\cdot 0.26)}\,\times 10^{-4}\,\text{GeV}^6$
that is in mutual relation with the pair $(b_2^\text{BMS},b_4^\text{BMS})$,
while the best fit of data processing in \cite{Mikhailov:2021znq} with $\chi^2_\text{ndf}\simeq 0.4$ provides $(b_2^{bf}(\mu^2_0)=\bm{0.112}, b_4^{bf}(\mu^2_0)= \bm{-0.029})$  for DA$^{bf}$.

 Note that both the BMSmod, see item \textbf{(i)}, and the platykurtic DA in item \textbf{(iii)} are included in the 1$\sigma$
level of the best-fit point DA$^{bf}$ in item \textbf{(iv)}, see Fig. \ref{fig:fitTFF-NNLO}.
The modern lattice results for the profile of the pion DA in \cite{Baker:2024zcd} demonstrate the same behavior
as the mentioned DA$^{bf}$ within $0.2 \leqslant x\leqslant 0.8$, see the corresponding profiles in Fig.\ref{fig:DAmodels}.

Summarizing, we will use those of nonperturbative inputs \textbf{(i-ii,iv)} that have already provided a good agreement at the
1$\sigma$ level between the experimental data of the process $\gamma+\gamma^* \to \pi^{0}$ and the corresponding predictions obtained within the theoretical framework of \cite{Mikhailov:2021znq} (see Fig.1 there) and \cite{Mikhailov:2021cee}.

\subsection{``Hybrid'' approach and preliminary comparisons} 
 \label{subsec:standard}
  \vspace{-3mm}

\begin{figure}[hbt]
 \includegraphics[width=0.55\textwidth]{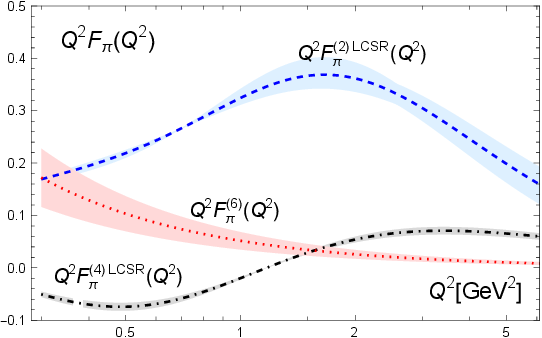}
 \vspace{-2mm} \caption{\footnotesize{\label{fig:Tw-2-4-6}
 Different twist contributions of the LO RG summation: the upper widening, blue strip is twist-2 $F^\text{(tw2)\text{LCSR,LO}}_\pi$;
the thin, dashed-dotted grey line is twist-4 $F^\text{(tw4)LCSR}_\pi$; the red narrowing strip with dots is 
twist-6 $F^\text{(tw6)}_\pi$ .}}
\end{figure}

Here we consider the results for the components of the  emFF:$F^\text{(tw2)LCSR,LO}_\pi$ -- for the leading twist-2 in Eq.(\ref{eq:LO-LCSR}) obtained within the LCSR with RG summation that are taken for the BMSmod DA \cite{Mikhailov:2021znq};
 $F^\text{(tw4)\text{LCSR}}_\pi$ -- for twist-4 in Eq.(\ref{eq:F4b}) obtained at the same condition;$F^\text{(tw6)}_\pi$ -- for the twist-6 in Eq.(\ref{eq:tw-6}).
The curves for all the components  are presented in Fig.\ref{fig:Tw-2-4-6} for comparison.
It is seen that the hierarchy of twist contributions is disturbed around small $Q^2_\text{thr}\approx 0.4$\,GeV$^2$.

\begin{figure}[htb]
\includegraphics[width=0.47\textwidth]{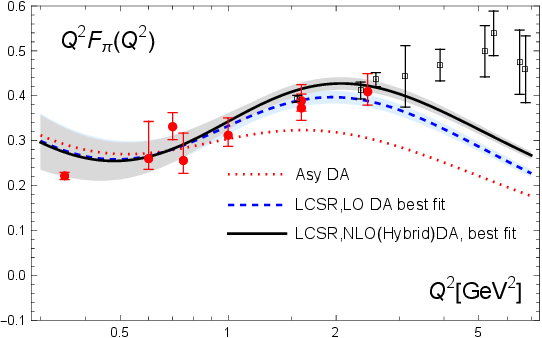}
 \includegraphics[width=0.47\textwidth]{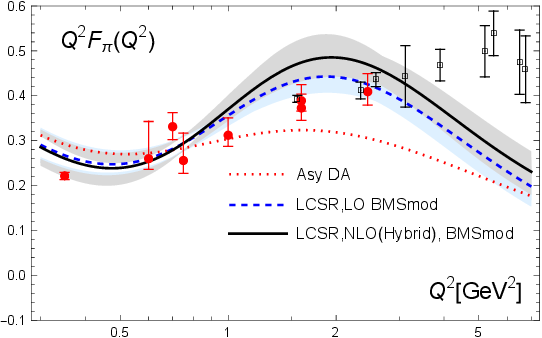}
\vspace{-3mm} \caption{\footnotesize{\label{fig:EMFF-RG}
Predictions for LO LCSR, $F^{\text{LCSR}}_{\pi} $ in Eq.(\ref{eq:F-RGtot}) for
DA$^{bf}$, for the bunch of BMSmod, and for Asy DAs are presented with dotted red line,
blue and grey strips - uncertainties in LO and NLO
respectively in both panels.
The red discs with error bars are the experimental data from \cite{PhysRevC.78.045203};
the open boxes with bars are the recent lattice predictions \cite{Ding:2024lfj}.
\textbf{Left:}Dashed blue line,
 and in addition to them the standard NLO LCSR \cite{Bijnens:2002mg} corrections -- upper solid black line (named hybrid) are
 based on DA$^{bf}$ from item \textbf{(iv)}.
 \textbf{Right:} The same  curve designations for $F^{\text{LCSR}}_{\pi}$ and NLO LCSR corrections based on the
 bunch of BMSmod DAs.
 }}
\end{figure}

Collected together, these components constitute  the self-consistent and completed result $F^{\text{LCSR}}_{\pi}$,
\be \label{eq:F-RGtot}
F^{\text{LCSR}}_{\pi}(Q^2)  = F^\text{(tw2)LCSR,LO}_\pi(Q^2)+F^\text{(tw4)\text{LCSR}}_\pi(Q^2)+F^\text{(tw6)}_\pi(Q^2)\,.
\ee
This $F^{\text{LCSR}}_{\pi}$ is  valuable and can be compared with the experimental data   independently,
which are presented in Fig.\ref{fig:EMFF-RG}.
There, we use the DA$^{bf}$ for the twist-2 in the left panel and the BMSmod DA in the right one at the parameters twist-4 and twist-6 from items \textbf{(ii)} and \textbf{(iv)} respectively.
The upper solid curves in both panels show ``hybrid'' estimates, for which  $F^{\text{LCSR}}_{\pi}$  in Eq.(\ref{eq:F-RGtot})
(dashed lines) are supplemented with the NLO corrections taken within FOPT from \cite{Bijnens:2002mg}.
Of course, these ``hybrid'' results appear at the mechanistic addition of NLO FOPT corrections
to the base that was summed by RG, i.e.,
they are the components of substantially different species.
Considering that the FOPT terms are always larger than the FAPT ones,  the upper solid curves
can be considered \textit{only as an upper bound} for the emFF.

One can conclude that both the pion DA models DA$^{bf}$ and bunch
of BMSmod DA \cite{Mikhailov:2021znq} (see Appendix \ref{App:E}) used sufficiently
well describe the experimental data on $F_\pi$ \cite{PhysRevC.78.045203},
if one takes into account inherent uncertainties of nonperturbative inputs.
In this regard, we note that  the larger strip of uncertainty in Fig.\ref{fig:EMFF-RG}(Right)
is due to the smearing of the whole bunch of BMS DAs.
At this stage of the analysis, one cannot reliably prefer one of these DAs to the other.
 \subsection{Estimate of the NLO contribution within FAPT}
 \label{subsec:lowerbound}
\begin{figure}[htb]
\includegraphics[width=0.47\textwidth]{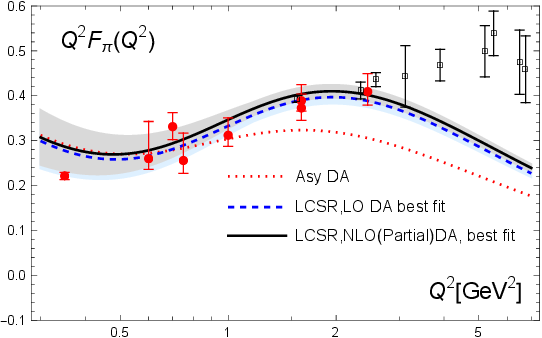}
 \includegraphics[width=0.47\textwidth]{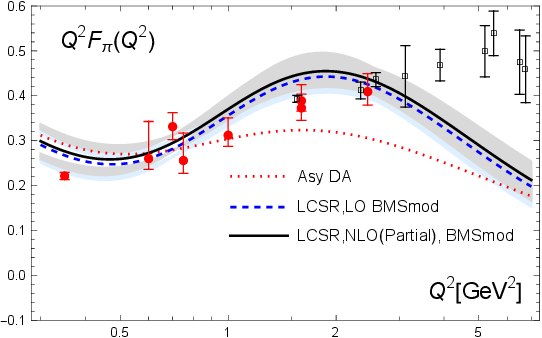}
\vspace{-3mm} \caption{\footnotesize{\label{fig:EMFF-FAPT}
The predictions for the (partial N)LO $F^{\text{LCSR}}_{\pi}$,
the  notations are the same as in Fig.\ref{fig:EMFF-RG} in both panels.
The red discs with their error bars -- experimental data from \cite{PhysRevC.78.045203};
the open boxes with bars --
 recent lattice results \cite{Ding:2024lfj}.
\textbf{Left:} Based on DA$^{bf}$ (item \textbf{(iv)}), dashed blue line presents LO,
upper solid black line, named ``Partial'', presents the (partial N)LO corrections in Eq.(\ref{eq:F_1aLCSR}).
 \textbf{Right:}The same designations are for the LO and the (partial N)LO to emFF, based on the  bunch of BMSmod DAs (item \textbf{(i)}). }}
\end{figure}

We have obtained the contribution to $H^{(1)}_{(n)}$ from the  pair of diagrams $(a,a')$ in the Feynman gauge,
\ba \label{eq:26}
H^{(1)}_{(n)} \Rightarrow H^{(1a)}_{(n)} =   H^\text{(0)}(u)
  \underset{u}{\otimes}
  \left[\1+ \bar{a}_s(u)\mathcal{H}^{(1a)}(u,v)\right]
   \left(\frac{\bar{a}_s(u)}{a_s(\mu^2_0)} \right)^{\nu_n}
   \underset{v}\otimes
   \psi_n(v) \,.
\ea
 The explicit expression for $T^\text{(1)}_{a}$ is presented in (\ref{eq:A1}) and contains the evolutional log,
 $L(z)$ accompanied by the $V^{a}$ part of the ERBL kernel, as was discussed in Sec.\ref{sec:theor-basis}.
 Refined $\mathcal{H}^{(1a)}$ presented in (\ref{eq:A3}) reads
\ba \label{eq:27}
\bar{a}_s(u)\mathcal{H}^{(1a)}(u,v)&=&\bar{a}_s(u)\, C_{\rm F}\,
U(u,v)\,.
\ea
It should be emphasized that we consider only a part of the NLO contribution related to the
``handbag'' diagrams here.
Substituting (\ref{eq:27}) in (\ref{eq:26}),
performing the replacements that are due to the dispersion relation
$$\bar{a}^{\nu }_s(u) \to \Delta_{\nu }(s(u),q(u)),~
\underset{v}\otimes \psi_n(v) \to \underset{v}\otimes (\theta(v \geqslant v_0)\psi_n(v)),~ \1=\delta(u-v),$$
 and using the notation $\Delta_{\nu}(u)\equiv \Delta_{\nu}(s(u),q(u))$,  one arrives to the result of the dispersion
relation under $H^{(1a)}_{(n)}$
\ba
H_{(n)}^\text{(1a)LCSR}&=&\frac{1}{a_s^{\nu_n}(\mu^2_0)}H_0(u)\underset{u}{\otimes} \bigg\{\Delta_{\nu_n}(u)\1 +\Delta_{(1+\nu_n)}(u)\,C_{\rm F}\, U(u,v)\Big]\bigg\}\underset{v}\otimes\!(\theta(v \geqslant v_0)\psi_n(v))\,. \label{eq:Hn2}
\ea
Now, we need to perform the Borel transformation under $H_{(n)}^\text{(1a)LCSR}$ in (\ref{eq:Hn2}),
as was done in Sec.\ref{sec:borel}. Recall  that
\ba
 M^2 \mathbf{\hat{B}}\Big[H_0(u)\Delta_{\nu_n}(u)\Big] &=&\left[\M_{\nu_n}\left(s(u)\right)\!+\int^{s(u)}_0\!\!\!\!\rho_{\nu_n}(s) \omega_1(s,u)ds\right] \exp{\Ds \left(\!-\frac{Q^2}{M^2}\frac{\bar{u}}{u}\right)}\,. \label{eq:borel-Delta}
\ea
So, using Eq.(\ref{eq:borel-Delta}) for the transformation of (\ref{eq:Hn2}), one obtains
\ba
\!\!\!\!\!\!\!\!F^\text{(tw2)LCSR,1a}_{\pi,n}(Q^2,M^2)&\!=\!&\! M^2 \mathbf{\hat{B}}H_{(n)}^\text{(1a)LCSR}
 \! =\!\frac{1}{a_s^{\nu_n}(\mu^2_0)}\Bigg\{
 \!\left[\M_{\nu_n}\left(s(u)\right)\!+\int^{s(u)}_0\!\!\!\!\rho_{\nu_n}(s)\omega_1(s,u)ds\right]\!\underset{u}\otimes \1 +
  \nonumber \\
  &&\!\!\! C_{\rm F} \left[\M_{(\nu_n+1)}(s(u))+\!\int^{s(u)}_0\!\!\!\!\rho_{(\nu_n+1)}(s) \omega_1(s,u)ds\right]\!\underset{u}\otimes\,U(u,v)\Bigg\}
  \exp{\Ds \left(\!-\frac{Q^2}{M^2}\frac{\bar{u}}{u}\right)} \nonumber \\
&&\underset{v}\otimes \left(\theta(v \geqslant v_0)\psi_n(v)\right)\,.  \label{eq:F_1aLCSR}
\ea
We obtain a NLO estimate of
$F^{\text{LCSR}}_{\pi}(Q^2)$ in Eq.(\ref{eq:F-RGtot})
replacing the LO $F^\text{(tw2)LCSR,LO}_{\pi,n}$ with the partial NLO result $F^\text{(tw2)LCSR,1a}_{\pi,n}$.
We will consider this result to be \textit{a lower bound} of the emFF in the NLO.
The final results at different nonperturbative inputs are presented in Fig.\ref{fig:EMFF-FAPT} with the same notations as in Fig.\ref{fig:EMFF-RG}.

The predictions presented in both panels of Fig.\ref{fig:EMFF-FAPT},
for DA$^{bf}$ (left) and for the bunch of BMSmod (right), respectively, look to be good.
But the curves with DA$^{bf}$ in Fig.\ref{fig:EMFF-FAPT}(left) look a bit more corresponding to the experimental data  in \cite{PhysRevC.78.045203}.
In addition, the estimate of the upper limit (hybrid) for this case, see Fig.\ref{fig:EMFF-RG}(left), does not change this correspondence.
\section{Conclusions}
\label{sec:concl}
We reconsider the perturbative part of the LCSR for the electromagnetic pion form factor (emFF) by
including renormalization group logarithm summation.
Together with the dispersion approach inherent in the LCSR, this summation necessarily leads
to fractional Analytic Perturbation Theory (FAPT) at the calculation of the emFF.
This approach is sequentially applied to twist-2 and twist-4 contributions.
It makes the size of radiative corrections effectively smaller and extends the
domain of applicability  of pQCD to the lower transfer momentum $Q^2 \sim 0.5$\,GeV$^2$.

The results for the pion emFF  are the effect of synthesis of FAPT,
 manifested within the  LCSR,
 and of nonperturbative inputs that are shaped by the intrinsic pion structure in the form of its distribution amplitudes of twist-2, $\varphi^{(2)}_\pi(x)$, twist-4, $\varphi^{(4)}_\pi(x)$, and twist-6.
These nonperturbative characteristics, $\varphi^{(2)}_\pi$ and $\delta^2_\text{tw-4}$,  were established within nonlocal
condensate sum rules approach in \cite{Mikhailov:1988nz,Bakulev:2001pa} and \cite{Bakulev:2002uc} respectively,
and confirmed within the LCSR analysis of the  pion transition form factor in
\cite{Ayala:2019etj,Mikhailov:2021znq}.

Our calculation of the emFF shows a good agreement with the experimental data \cite{PhysRevC.78.045203},
see Fig.\ref{fig:EMFF-FAPT},
in spite of the incompleteness of the NLO of the pQCD corrections.
The completed NLO calculation for the emFF will allow one to perform the numerical analysis and reliably extract
from the experimental data the pion distribution amplitude $\varphi^{(2)}_\pi$ of the leading twist.

\acknowledgments
This work was supported in part by FONDECYT (Chile) Grant No.~1240329 (C.A.).
S.V.M. is very thankful to Universidad T\'ecnica Federico Santa Mar\'ia,  Valpara\'iso, Chile, where this work was started,
  for hospitality.

\begin{appendix}
\appendix
\section{QCD perturbative expansion beyond the leading order}
\label{App:A}
The factorized amplitudes $T^\text{(1)}$ of the partonic subprocess for $\pi-\gamma$, see Eq.(\ref{eq:3b}), that correspond to the ``handbag'' diagrams $(a,a')$ taken in Feynman gauge and presented, \eg, in the first row of Fig.3 in \cite{Braun:1999uj}
($\mu_F^2=\mu_R^2=\mu^2$),  reads \\
\begin{subequations}
 \label{eq:A1}
 \begin{eqnarray}
a_s(\mu^2)T^\text{(1)}_a(y) &=&a_s(\mu^2) C_{\rm F}\, H_0(x) \underset{x}{\otimes}\left\{ L(x)\,2 V^{a}(x,y)  + \left[\frac{\theta(\bar{y}>\bar{x})}{\bar{y}} - V^{a}(x,y)\right] \right\}; \label{eq:A1a}   \\
a_s(\mu^2) T^\text{(1)}_{a'}(y) &=&a_s(\mu^2) C_{\rm F}\, H_0(x) \underset{x}{\otimes}\left[L(x)-1 \right]\,\1(x,y); \label{eq:A1b} \\
&&L(x)=\ln\left( \frac{q(x)}{\mu^2}\right);\,q(x) = \sigma x + Q^2\bar{x}\,. \label{eq:A1d}
\end{eqnarray}
 \end{subequations}
The first term in braces in (\ref{eq:A1a}) corresponds to the  ERBL-evolution logarithm discussed before in Eq.(\ref{eq:Fn-1loop});
this term is transferred later to the common factor with $\left(\bar{a}_s(u)\right)^\nu$ in Eq.(\ref{eq:Fn-1loop}).
The second term (in square brackets) shapes the term $\bar{a}_s(u) \mathcal{H}^{(1)}(u,v)$  in Eq.(\ref{eq:Fn-1loop}).
The logarithm $L(x)$ in (\ref{eq:A1b}) (diagram $a'$) is of ultraviolet nature ($\mu=\mu_R$) and cancels among other diagrams
(including renormalization of quark legs) due to the Ward identity.
The other diagrams in Fig.3 \cite{Braun:1999uj}  should lead to the contributions related to the part $V^{b}$
of the whole  kernel $V_0$ and  some remainder.
The elements $V^{a},\,V^{b}$ of the one-loop evolution kernel $V_0$ are
\begin{subequations}
\begin{eqnarray} 
 V_0(x,y)  &=&C_{\rm F} V^{(0)}_+(x,y)\equiv C_{\rm F} 2 \left[V^{a}(x,y) +V^{b}(x,y)\right]_{+} =
C_{\rm F} 2\left[{\cal C}\theta(y>x)\frac{x}{y}
         \left(1+\frac{1}{y-x}\right)
         \right]_{+}\,; \label{eq:V}\\
V^{a}(x,y) &=&
   {\cal C} \theta(y>x)\frac{x}{y};~~
V^{b}(x,y)=
  {\cal C} \theta(y>x)\frac{x}{y}\left(\frac{1}{y-x}\right)\, ,
\end{eqnarray}
 \begin{eqnarray}
 V_0(x,y)\otimes\psi_n(y)  &=& C_{\rm F} 2 \left[V^{a}(x,y) +V^{b}(x,y)\right]_{+}\otimes\psi_n(y)=-\frac1{2} \gamma_{0}(n)\,\psi_n(x)\,,
\end{eqnarray}
\end{subequations}
where the symbol $\mathcal{C}$ means
$ {\cal C}=1+\left\{ x \to \bar{x}, y \to \bar{y}\right\} $.
Based on (\ref{eq:A1}) one obtains for the kernel $\mathcal{H}^{(1)}$ 
\ba \label{eq:A3}
         \mathcal{H}^{(1)}_a(u,v) &=& C_{\rm F}\, U(u,v),\, \text{where}
         ~U(u,v)=\frac{\theta(\bar{v}>\bar{u})}{\bar{v}}-V^{a}(u,v)\,.
\ea
The leading-order coefficient of the $\beta$ function,
 and the anomalous dimension $\gamma_{}(n)$ used in the above equations are
 \begin{subequations}
\label{eqA:RG-1l}
\ba
 \frac{d}{dL}\bar{a}_s &=& -\beta\left(\bar{a}_s\right)=-\bar{a}_s^2\left(\beta_0+\bar{a}_s \beta_1+\ldots \right)\,, \label{eqA:RG-1la}\\
  \beta_0&=&
  \frac{11}{3}{\rm C_A} - \frac{4}3 T_{\rm R} n_f;~
  \gamma_{0}(n)=2 C_\text{F}\left[ 4\left(\psi(2+n)-\psi(2) \right) -\frac{2}{(n+1)(n+2)}+1 \right];~ \nu_n =-\frac1{2} \frac{\gamma_{0}(n)}{\beta_0},
\label{eqA:RG-1lb}
\ea
 \end{subequations}
with $n_f$ being the number of active flavors ($n_f=3$ here) and
$T_{\rm R}=1/2, {\rm C_F}=4/3, {\rm C_A}=N_c=3$ for SU$_c$(3).
\section{FAPT coupling constants and their properties}
 \label{App:B}
In this Appendix, we give the expressions for the standard one-loop
running couplings and their FAPT counterparts.
To facilitate the representation, we express them in terms of the
auxiliary variables
$L=\ln(Q^2/\Lambda_\text{QCD}^2), L_s=\ln(s/\Lambda_\text{QCD}^2)$,
multiplied for simplicity by the term $\beta_0^\nu$.
In other words, we shift the origin of the different coupling images
to the values $a_s \to A_s= \beta_0 a_s =\beta_0 \alpha_s/(4\pi)$,
the corresponding auxiliary FAPT coupling constants are shown with the symbol ``bar",
$\bar{{\cal A}}_\nu=\beta_0^\nu\, \A_\nu$, $\bar{{\mathfrak A}}_{\nu}=\beta_0^\nu\, \M_{\nu},\, \bar{\I}_{\nu}=\beta_0^\nu\, \I_{\nu}, \bar{\rho}_{\nu}=\beta_0^\nu\, \rho_{\nu}$, in this section.
\subsection{The ``minimal'' FAPT coupling constants ${\cal A}_{\nu},  {\mathfrak A}_{\nu}$ at one-loop running}
The properties of the ``minimal'' FAPT couplings constants at one-loop running with $L$ and $L_s$ reads,
\begin{subequations}
\label{eq:couplings}
\begin{eqnarray}
A_{s}^{\nu}[L] &=& \frac1{L^\nu}~~~~~~~~~~~~~~~~~~~~~~~~~~~~~~~~~~~~~~~~~~~~~~~~~~~~~~~~~~~~~~~~~~~~~~~~~~~~~~~~~~~~~~~ \mbox{standard pQCD}\, ,
\label{eq:A-s} \\
 \bar{{\cal A}}_{\nu}[L]
          &=& \frac1{L^{\nu}}
   - \frac{F(e^{-L},1-\nu)}{\Gamma(\nu)}\,; ~~~~~~~~~~~~~~~~~~~~~~~~~~ \bar{{\cal A}}^{}_{1}[L]
 =\frac1{L}- \frac1{e^L -1}\,,~~~~~~~~~~~~~~~\mbox{spacelike FAPT}\, ,\label{eq:A-F} 
 \end{eqnarray}
 \begin{eqnarray}
  \bar{{\cal A}}_{0}[L]=1;&~ \bar{{\cal A}}_{0<\nu <1}[L \to -\infty] \rightarrow |L|^{1-\nu};~
 \bar{{\cal A}}_{1}[L \to -\infty]\to 1;&~ \bar{{\cal A}}^{}_{\nu >1}[L \to -\infty]\to 0\, ,
\label{eq:B1c}
\end{eqnarray}
 \begin{eqnarray}
 \bar{{\mathfrak A}}^{}_{\nu}[L_s]
  & = & \frac{\sin\left[(\nu -1)\phi \right]}
             {\pi\,(\nu -1) \left(L_s^2+\pi ^2\right)^{(\nu-1)/2}}\,;~~
        \bar{{\mathfrak A}}^{}_{1}[L_s]=\frac{\phi}{\pi};~~\phi=\arccos\left(\frac{L_{s}}{\sqrt{L_{s}^2+\pi^2}}\right),~~ \mbox{timelike FAPT} ,
\label{eq:U-F}
\end{eqnarray}
 \begin{eqnarray}
 \bar{{\mathfrak A}}^{}_{0}[L_s]=1;&~ \bar{{\mathfrak A}}^{}_{0<\nu <1}[L_s \to -\infty]
\rightarrow \left(\sqrt{L_s^2+\pi^2}\right)^{1-\nu};&~ \bar{{\mathfrak A}}^{}_{1}[L_s
\to -\infty]\to 1;~ \bar{{\mathfrak A}}^{}_{\nu >1}[L_s \to -\infty]\to 0\,.
\label{eq:B1e}
\end{eqnarray}
 \end{subequations}
where the symbol $[L]$ is used to denote the function argument,
clearly distinguishing it from the $Q^2$ dependence.\footnote{An
expression analogous to (\ref{eq:U-F}) was derived long ago in
\cite{Gorishnii:1983cu,Broadhurst:2000yc} in connection with multiloop
calculations and the Higgs-boson decay into hadrons.}
In Eq.(\ref{eq:A-F}) there appears the ``pole remover'' $\Ds \frac{F(e^{-L},1-\nu)}{\Gamma(\nu)}$ -- to the standard $A_{s}^{\nu}(L)$,
it is expressed in terms of the Lerch transcendental function, or is $\Ds \frac{\Li{(1-\nu)}(e^{-L})}{\Gamma(\nu)}$,
$F(z,\nu)=\Li{\nu}(z)$ \cite{Bakulev:2005gw}. Below we show the $Q^2$-behavior of the first $\left(\bar{{\cal A}}_{1}(Q^2),\bar{{\mathfrak A}}^{}_{1}(s)\right)$ and second $\left(\bar{{\cal A}}_{2}(Q^2),\bar{{\mathfrak A}}^{}_{2}(s)\right)$ analytic couplings for illustration

\includegraphics[width=0.45\textwidth]{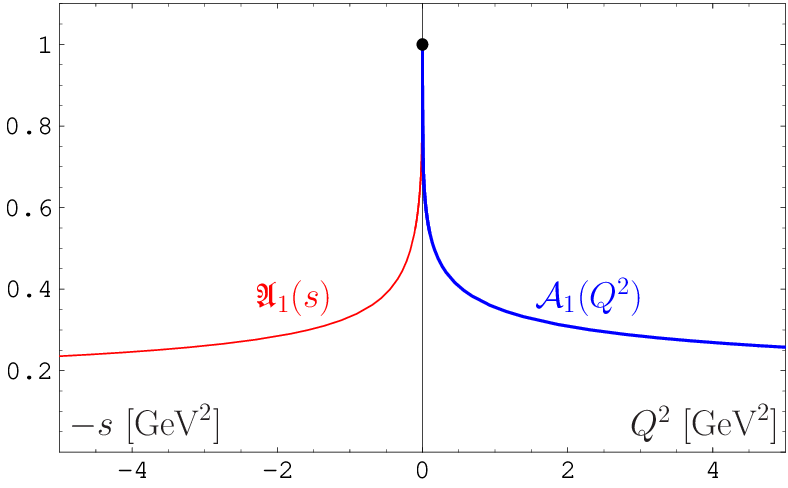}~
 \includegraphics[width=0.46\textwidth]{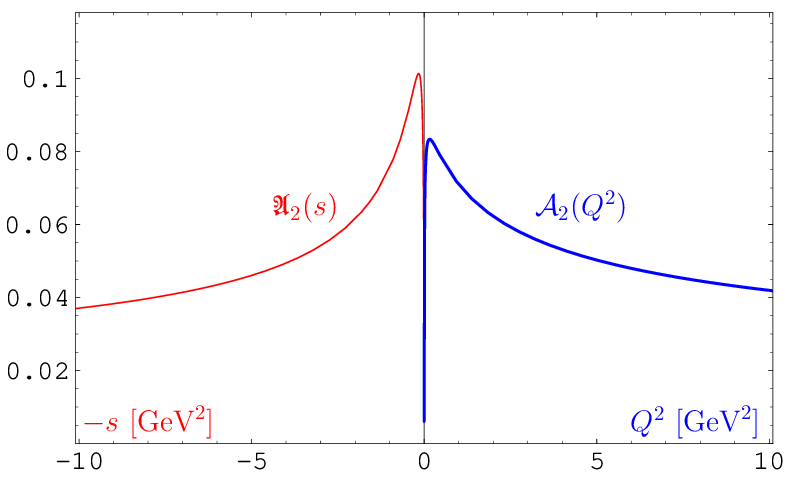}

It is seen from (\ref{eq:B1c},\ref{eq:B1e}) that the couplings
$ \bar{{\cal A}}^{}_{ \nu }[L], ~ \bar{{\mathfrak A}}^{}_{ \nu }[L_s]$
become unbounded in the vicinity of $Q^2=0$ and $0< \nu \leqslant 1$.
It can be interpreted as that  the well-known singularity of the standard
running coupling $A^{\nu}_{s}[L]$ in Eq.\ (\ref{eq:A-s}) at
$L=0 (Q^2=\Lambda^2)$ and $\nu> 0 $ turns
into a singularity of the FAPT couplings in the limit $L \to -\infty$ for
$0< \nu <1$, cf.\ (\ref{eq:B1c}) and (\ref{eq:B1e}).
Therefore, these results for
$ \bar{{\cal A}}^{}_{\nu}[L], ~ \bar{{\mathfrak A}}^{}_{\nu}[L_s]$
cannot be directly used in the vicinity of origin at $0< \nu \leqslant 1$,
where the FAPT couplings are ill-defined.
We redefine these couplings in this regime
by demanding that they vanish together with its first derivatives.
This intervention guarantees, among other things,  that observables
calculated with them, e.g., the pion form factors \cite{Ayala:2018ifo,Mikhailov:2021znq,Mikhailov:2021cee}, have the correct UV asymptotics
following from pQCD.
To this end, we redefine the couplings
$ \bar{{\cal A}}_{\nu},~ \bar{{\mathfrak A}}_{\nu}$
in the vicinity of the origin
\be \label{eq:A-constraint}
\text{and, for $0< \nu \leqslant 1$:}~  \bar{{\cal A}}_{\nu}[-\infty]=0,\, \bar{{\cal A}}_{\nu}'[-\infty]=0,~ \bar{{\mathfrak A}}_{\nu}[-\infty]=0,\, \bar{{\mathfrak A}}_{\nu}'[-\infty]=0\,,
\ee
while the behavior of these couplings for $\nu > 1$ remains unaffected.
It should be mentioned that  the constraints like (\ref{eq:A-constraint}) can be realized in the modeling of low-energy behavior
following the way  \cite{Ayala:2016zrz,Ayala:2017tco}.
We prefer to avoid the models  here; instead, we formulate these  constraints at the point $Q^2=0$ explicitly.
Simultaneously, we hope that the integration over this vicinity does not contribute significantly.
\subsection{Spectral density $\rho_\nu$ and coupling $\I_{\nu}$}
The auxiliary spectral density $\bar{\rho}^{}_\nu$ has the form
($L_\sigma=\ln(\sigma/\Lambda_\text{QCD}^2)$),
\begin{equation}
 \bar{\rho}_{\nu}^{}(\sigma)=  \bar{\rho}_{\nu}^{}[L_\sigma]\equiv \beta_0^\nu \rho_\nu[L_\sigma] = \frac{1}{\pi}\textbf{Im}\,\big[A^{\nu}(-\sigma)\big]
\stackrel{\text{1-loop}}{\longrightarrow} \frac{1}{\pi}\,
  \frac{\sin\left[\nu~\phi\right] }{\left(L^2_\sigma+\pi^2\right)^{\nu/2}} \,,~~~\bar{\rho}_{1}^{}[L_\sigma]= \frac{1}{L^2_\sigma+\pi^2}\,,
\label{eq:spectr-dens}
\end{equation}
see the definitions of $A^{\nu}$ in (\ref{eq:A-s}) and $\rho_\nu$ in (\ref{eq:key_disper}).
The $ \bar{\rho}_{\nu}^{}(\sigma)$ is the origin of FAPT coupling constants
\ba
\bar{\A}_{\nu}(x)=\int_0^{\infty} \frac{d\sigma}{\sigma+x}\bar{\rho}_{\nu}^{}(\sigma) &,& ~\bar{\M}_{\nu}(y)=\int_y^{\infty}\frac{d\sigma}{\sigma}\bar{\rho}_{\nu}^{}(\sigma)\,,
\label{eq:faptcouplings}
\ea
presented in the one-loop approximation in Eqs.(\ref{eq:A-F},\ref{eq:U-F}).
 All of the invented coupling constants, $\{A_{s}^{\nu},   \bar{{\cal A}}_{\nu},  \bar{{\mathfrak A}}_{\nu}, \bar{\rho}_{\nu} \}$ satisfy the one-loop RG equation that we show below for the partial case of $\bar{\rho}_\nu$ 
\be \label{eq:RG-1loop}
 \frac{d}{dL}\bar{\rho}_{\nu}[L] = - \nu \,\bar{\rho}_{\nu+1}[L]\,.
\ee

The generalized FAPT coupling $ \I_{\nu}$ is the function of two arguments; it appears  due to the action of  ``subtraction of continuum'' \footnote{It should be mentioned that $\I_{1}(y,x)$ also appeared in \cite{Ayala:2017tco,Ayala:2024ghk} due to the modeling of spectral density $\rho_1(s)$ at  low energy.}
 and reads
\begin{eqnarray}
 \label{eq:B7}
   \bar{\I}_{\nu}(y,x)=
\int_{y}^\infty \frac{ds}{s+x} \bar{\rho}_{\nu}(s)
&=&\left\{
\begin{array}{l}
 \Ds
  \left[ \bar{\M}_{\nu}(y) -~~ x\int_{y}^\infty \frac{ds}{s(s+x)} \bar{\rho}_{\nu}(s) \right]
  \leqslant \bar{\M}_{\nu}(y)\, (\text{for}~ \bar{\rho}_{\nu}\geqslant 0),
\label{eq:A-def} \\
\Ds \left[ \bar{\A}_{\nu}(x)\! -\! \int_{0}^y \frac{ds}{s+x} \bar{\rho}_{\nu}(s) \right]
   \leqslant\bar{\A}_{\nu}(x)\,(\text{for}~ \bar{\rho}_{\nu}\geqslant 0),
 \end{array}
 \right. \\
   \bar{\I}_{\nu}(y, 0)
&=& \bar{\M}_{\nu}(y),
~~ \bar{\I}_{\nu}(0, x)
= \bar{\A}_{\nu}(x),
~~ \bar{\I}_{1}( 0, 0)=   \bar{\M}_{1}(0)= \bar{\A}_{1}(0)\,.
\label{eq:B8}
\end{eqnarray}
The coupling $\I_{\nu}(y,x)$ is regular for $y>0, x>0$,
while for $y=0 $ or $x=0$ it reduces to the initial FAPT couplings
in accordance with Eq.\ (\ref{eq:B8}).
For the important case $\nu=1$, the expression for $ \bar{\I}_{1}(y,x)$ reduces to
\be
  \bar{\I}_{1}(y,x)=\frac1{L_x}- \frac1{e^{L_x} -1} - \int_{0}^y \frac{ds}{s+x}\left(\frac{1}{L^2_s+\pi^2}\right),
\ee
where $L_x =\ln(x/\Lambda^2)$, e.g.,
\be
 \bar{\A}_{1}(\Lambda^2)= \bar{\M}_{1}(\Lambda^2)=\frac{1}{2}>  \bar{\I}_{1}(\Lambda^2,\Lambda^2)
\simeq  \bar{\A}_{1}(\Lambda^2)-0.06\thickapprox
 \bar{\A}_{1}(\Lambda^2)-\left(\frac{\ln(2)}{\pi^2}- \frac{3}{4}\frac{\zeta_3}{\pi^4}\right)
\ee
\subsection{Extended boundary conditions}
 \label{subsec:boundary}
For the twist-4 contribution, see Eq.(\ref{eq:F4b}),
we face with a new effective coupling constant $ \bar{\I}'_{\nu}(s(u),0)$
\begin{subequations}
  \label{eq:B9}
\ba \label{eq:B9a}
  \bar{\I}'_{\nu}\left(s(u),0\right)\equiv \frac{d}{dx} \bar{\I}_{\nu}(y,x)\Big{|}_{x=0}= -\int_{s(u)}^{\infty} \frac{ds}{s^2}\bar{\rho}_\nu(s)\,,
 \ea
 that enters later into the convolution with $\varphi(u)=\varphi^{(4)}_{\pi}(u) e^{\left(\!-\frac{Q^2}{M^2}\frac{\bar{u}}{u}\right)}\,/u$.
The application to the last term in (\ref{eq:B9a}) sequentially integrating by parts and the usage of (\ref{eq:RG-1loop}) leads to the representation
\ba \label{eq:B10}
\!\!\!\!\int_{s(u)}^{\infty} \frac{ds}{s^2}\bar{\rho}_\nu(s)&\!=\!&\left(\frac{\bar{\rho}_{\nu}(s(u))}{s(u)}-\nu \frac{\bar{\rho}_{\nu+1}(s(u))}{s(u)}+\nu(\nu+1)\int_{s(u)}^{\infty} \frac{ds}{s^2}\bar{\rho}_{\nu+2}(s) \right)\!=\! \sum_{n=0} (-1)^n \frac{\Gamma(\nu+n)}{\Gamma(\nu)} \frac{\bar{\rho}_{\nu+n}(s(u))}{s(u)}, \\
\!\!\!\! \text{and} && \frac{\bar{\rho}_{\nu+n}(s)}{s}= \frac{e^{-\pi t}}{\pi^{(1+\nu+n)}} \frac{\sin\left[(\nu+n)\phi\right]}{(1+t^2)^{(\nu+n)/2}}\,,~ \phi=\arccos\left(\frac{t}{\sqrt{1+t^2}}\right)\,,
\ea
\end{subequations}
where $t= \ln(s/\Lambda^2)/\pi,\, s(u)=(s_0+Q^2) (u - u_0)$.
In other words, we expanded $ \bar{\I}'_{\nu}(s(u),0)$ in (\ref{eq:B9a}) in a series in the standard coupling constants
$\bar{\M}'_{\nu+n}(s(u))$.
Every term in the sum in the rhs of Eq.(\ref{eq:B10}),
starting with the second one at $n=1$, is integrable over $u$,
while the first term, $\Ds -\frac{\bar{\rho}_{\nu}(s)}{s}=\frac{d}{ds} \bar{\M}_{\nu}(s)\Big|_{s=s(u)}$, is nonintegrable \textit{per se}
due to the inclusion of the vicinity of $s(u_0)=0$.
Nevertheless, taking the corresponding integral in (20) by part and using the  condition (\ref{eq:A-constraint}) for the ``surface term'', one obtains
\be \label{eq:key-int}
  \int_{u_0}^1 \bar{\M}'_{\nu+n}(s(u))\,\varphi(u) du = 0 - \int_0^{s_0} \bar{\M}_{\nu+n}(s)\, \varphi'\left(\frac{Q^2+s}{Q^2+s_0}\right) \frac{ds}{(Q^2+s_0)^2}\,,
\ee
where the integral in the rhs is convergent even at $n=0$.
One can use  a few (two) first terms in the expansion (\ref{eq:B10}), which is enough for our NLO calculation accuracy.
The final expression for the integration of  $\bar{\I}'_{\nu}(s(u),0)$  based on (\ref{eq:key-int}) and as applied to  Eq.(\ref{eq:F4b}) reads
\ba \label{eq:final-key-int}
\!\!\int_{u_0}^{1}\!\!\I'_{\nu_{t4}}(s(u),0)\varphi(u) du =-\frac{1}{(\beta_0)^{\nu_{t4}}}\int^{s_0}_0\!\!\! \left( \bar{\M}_{\nu_{t4}}(s)
+ \sum_{n=1} (-1)^n \frac{\Gamma(\nu+n)}{\Gamma(\nu)} \bar{\M}_{(\nu_{t4}+n)}(s) \right)\! \varphi'\left(\frac{Q^2+s}{Q^2+s_0}\right) \frac{ds}{(Q^2+s_0)^2}
\ea
\section{Borel transformation}
 \label{App:C}
We used the standard form of the Borel transform for QCD SR, see \eg in \cite{Shifman:1978bx}, it reads $\hat{\mathbf{B}}_{\scriptstyle{(M^2 \to \sigma)}}[f(\sigma)]$
and manifests itself as the limit of a series of derivatives of the function $f(\sigma)$
\begin{gather}
\label{eq:borel}
\!\!\!\!\!\hat{\mathbf{B}}_{\scriptscriptstyle{(M^2 \to \sigma)}}[f(\sigma)]\!\equiv\!\mathbf{\hat{B}} \left[ f(\sigma) \right]\!\!(M^2)\!\stackrel{def}{=}\!\!\lim_{\begin{subarray}{c} \scriptscriptstyle{\sigma=n M^2} \\ \scriptscriptstyle{n\to \infty}\end{subarray}} \frac{(-\sigma)^n}{\Gamma(n)} \frac{\mathrm d^n }{\mathrm d \sigma^n} \left[f(\sigma)\right].
\end{gather}
Further, for shortness, we will omit the subscript at $\hat{\mathbf{B}}_{\scriptscriptstyle{(M^2 \to \sigma)}} \equiv \hat{\mathbf{B}}$.
\begin{eqnarray} \label{eq:B2}
\!\!\!\!\mathbf{\hat{B}} \left[ \exp\left(- \sigma a\right)\right]\!&=&\! \delta\left(1- M^2 a\right).
\end{eqnarray}
 Based on (\ref{eq:B2}), one can derive 
\begin{eqnarray} \label{eq:B3}
 \mathbf{\hat{B}}\left[\left(\frac{1}{\sigma}\right)^\nu\right]&=&\frac{1}{\Gamma(\nu)}\left(\frac{1}{M^2}\right)^\nu,~
\mathbf{\hat{B}} \left(\frac{1}{\sigma+Q}\right)^\nu =  e^{\Ds -\frac{Q}{M^2}}\frac{1}{\Gamma(\nu)}\left(\frac{1}{M^2}\right)^\nu\,.
\end{eqnarray}
For the $\sigma$-dependence in (\ref{eq:F4a}), where $x = q(u)= \sigma \,u + Q^2\, \bar{u}\,$, one obtains
\begin{subequations}
\label{eq:B4}
\begin{eqnarray}
M^2 \mathbf{\hat{B}}\left[\frac{u}{ q(u)}\right] &=&e^{\Ds -\frac{Q^2}{M^2}\frac{\bar{u}}{u}}\label{eq:B4a};\\
M^2 \mathbf{\hat{B}}\left[\frac{u}{\left(s+q(u)\right) q(u)} \right] &=& \frac{1}{s}\left[1-\exp{\left(-\frac{s}{M^2 u}\right)}\right] e^{\Ds -\frac{Q^2}{M^2}\frac{\bar{u}}{u}} \label{eq:B4b};\\
M^2 \mathbf{\hat{B}}\left[\frac{u}{\left(s+q(u)\right) q(u)^2} \right] &=&
\frac{1}{s^2}\left[\exp{\left(-\frac{s}{M^2 u}\right)}-\left(1-\frac{s}{M^2 u}\right)\right] e^{\Ds -\frac{Q^2}{M^2}\frac{\bar{u}}{u}}\,.
\label{eq:B4c}
\end{eqnarray}
 \end{subequations}
The Borel transform of one-loop QCD running $\Ds \bar{a}_s(q(u))=\frac{1}{\beta_0 \ln(q(u)/\Lambda^2)}$,
\begin{subequations}
\label{eq:B5}
\begin{eqnarray}
M^2 \mathbf{\hat{B}}\left[\bar{a}_s^{\nu}(q(u))\right]&=&M^2\exp\left(-\frac{Q^2}{M^2}\frac{\bar{u}}{u}\right)\frac{\nu}{\beta_0^\nu} \bm{\mu} \left(\frac{\Lambda^2}{u M^2}, \nu \right)\,; \label{eq:B5a}\\
M^2 \mathbf{\hat{B}}\left[\bar{a}_s^{\nu}(q(u))\frac{u}{q(u)}\right]&=&\exp\left(-\frac{Q^2}{M^2}\frac{\bar{u}}{u}\right) \frac{1}{\beta_0^\nu}\bm{\mu} \left(\frac{\Lambda^2}{u M^2}, \nu-1 \right)\,;\label{eq:B5b} \\
\text{where} &&\bm{\mu}(t, \nu)=\int_0^\infty dx \frac{x^\nu}{\Gamma(\nu+1)}\frac{t^x}{\Gamma(x+1)}\,.\label{eq:B5c}
 \end{eqnarray}
  \end{subequations}
The transcendental function $\bm{\mu}(t,\nu)$ in (\ref{eq:B5}) appeared in similar calculations on the Borelization in
\cite{AlamKhan:2023ili}.
Now, let us introduce the Borel image $B(x,\nu)$ of the QCD one-loop running coupling $A_s^\nu(\sigma)$ in (\ref{eq:A-s}),
images $B$ are the elements of non-power perturbative series
\ba
  A_s^\nu \stackrel{\hat{B}}{\rightarrow} \mathbf{\hat{B}}\left[ A_s^\nu \right]=\nu \bm{\mu}\left(\frac{\Lambda^2}{M^2},\nu \right) \stackrel{def}{=} B\left(\frac{\Lambda^2}{M^2}=x,\nu \right)\,.
\ea
The RG equation for the image $B\left(x,\nu \right)$ reads
\be \label{eq:C7}
   x \frac{d}{d x}B\left(x,\nu \right)= - \nu B\left(x,\nu +1 \right),
\ee
that coincides (in the form) with the one-loop RG equation $\Ds \sigma \frac{d}{d\sigma} A^\nu(\sigma) = - \nu A^{\nu+1}(\sigma)$ for the standard power series.

The Borel images of $\bar{a}_s^{\nu}$ in Eqs.(\ref{eq:B5}-\ref{eq:C7}) are \textit{new results}. It should be mentioned that the behavior of these images in (\ref{eq:B5}) does not correspond to the behavior of $\bar{a}_s^{\nu}(\mu^2_{u})=\bar{a}_s^{\nu}(M^2 u+Q^2 \bar{u})$, where the ansatz
\be \label{eq:ansatz}
\mu^2 \to \mu^2_{u}=M^2 u+Q^2 \bar{u}
\ee
was suggested in \cite{Braun:1999uj} and used in \cite{Braun:1999uj,Bijnens:2002mg,Cheng:2020vwr}.

\section{Higher twist contributions}
\label{App:D}
 The explicit expressions for the twist-4 \cite{Bijnens:2002mg} amplitude $H^{(4)}_\pi$ read
\begin{eqnarray}
 \label{eq:Cx}
H^{(4)}_\pi(Q^2) = \int^{s_0}_0 \frac{ds'}{s'+\sigma} \rho_\text{tw-4}(s');~
\rho_\text{tw-4}(s')= \frac{\delta_\text{tw-4}^2(\mu^2)}{u^2}\frac{d}{ds'}\delta\left(\frac{\bar{u}}{u}\,Q^2-s'\right)\otimes
(u \varphi^{(4)}(u))
\end{eqnarray}

\ba
H^{(4)}_\pi(Q^2)& =& \delta_\text{tw-4}^2(\mu^2)\left[ \frac{\theta(u>u_0)}{\left(\Ds u\sigma+\bar{u}Q^2 \right)^2}+
                        \frac{\delta(u -u_0)}{ \left(\Ds u\sigma+\bar{u}Q^2 \right)}\, \frac{1}{s_0+Q^2}\right]
                        \otimes (u \varphi^{(4)}_\pi(u)) \label{eq:D2a}\\
                &=&\delta_\text{tw-4}^2(\mu^2)\left[\int_{u_0}^1 du \frac{\varphi^{(4)}_\pi(u)}{u\left( \sigma+\frac{\bar{u}}{u}Q^2 \right)^2}+
                \frac{\varphi^{(4)}_\pi(u_0)}{ \sigma+\frac{\bar{u}}{u} Q^2} \frac{u_0}{Q^2}\right]\,, \label{eq:D2b}
                        \ea
where $\Ds u_0=\frac{Q^2}{Q^2+s_0}, \bar{u}_0=\frac{s_0}{Q^2+s_0}$. Applying the Borel transform $M^2 \mathbf{\hat{B}}$ to the amplitude $H^{(4)}_\pi$, one obtains \cite{Bijnens:2002mg}
\ba \label{eq:finalF4}
F^{(4)}_\pi(Q^2,M^2) = M^2 \mathbf{\hat{B}}\left[H^{(4)}_\pi(Q^2,\sigma)\right] =
\delta_\text{tw-4}^2(\mu^2)\left[\int_{u_0}^1 du \frac{\varphi^{(4)}_\pi(u)}{u M^2} \Ds e^{-\frac{\bar{u}}{u}\frac{Q^2}{M^2}}+
                \varphi^{(4)}_\pi(u_0) \frac{u_0}{Q^2} e^{-\frac{s_0}{M^2}}\right]\,,
\ea
To estimate  $\delta_\text{tw-4}^2(\mu_0^2=1\,\text{GeV}^2)$ it was related to $\lambda^2_q$ using QCD SR approach in \cite{Bakulev:2002uc}.
The numerical estimate is $\delta_\text{tw-4}^2(\mu_0^2=1\,\text{GeV}^2) \approx \lambda^2_q/(2\cdot 1.075)=(0.4 - 0.45)/(2\cdot 1.075)\approx 0.198\pm 0.02$\,GeV$^2$.
The common factor $\delta_\text{tw-4}^2(\mu^2)$ should be transferred under the integral in the rhs of (\ref{eq:D2a}) when applying the ansatz $\mu^2=M^2 u+Q^2 \bar{u}$.
\section{Pion DAs of twist-2}
 \label{App:E}
\begin{figure*}[htb]
		\includegraphics[width=0.5\textwidth]{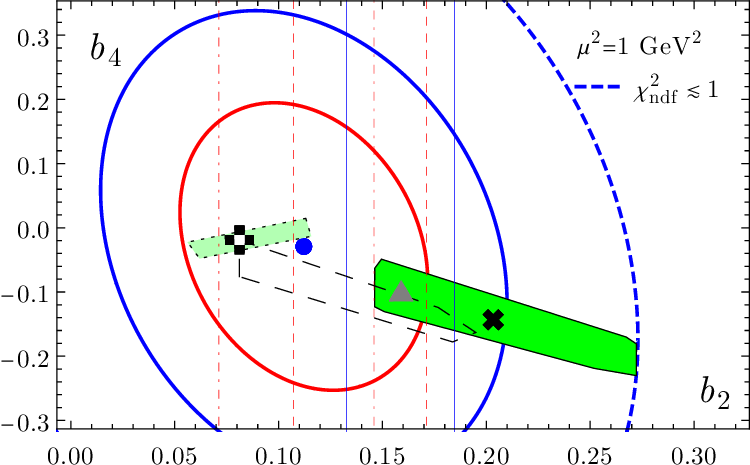}\vspace{-4mm}
	\caption{\label{fig:fitTFF-NNLO} (color in line) \footnotesize
	   The conformal expansion coefficients $b_2,b_4$ are on the axes.
		The following notations are used:
        black \ding{54} is BMS DA \cite{Bakulev:2001pa};
        black/white  \ding{60} is the platykurtic DA \cite{Stefanis:2014nla,Stefanis:2015qha};
		DAmod is shown as a grey \GrayW{\ding{115}}$\left(0.159^{+0.025}_{-0.027},-0.098^{+0.05}_{-0.03}\right)$
		selected from the BMS set of DAs to be inside the $1\sigma$ error ellipse (innermost red line),
		while vertical solid lines are the N$^3$LO
		lattice constraints on $b_2$ from lattice QCD \cite{Bali:2019dqc}.
			Results of the fitting procedure for the twist-two conformal coefficients
		$b_2,b_4$ with fixed higher-twist parameters is shown as blue best fit  \BluTn{\ding{108}} -- $(b_2^{bf}=0.112, b_4^{bf}= -0.029)$.
		Two rectangles along the lower diagonal denote the range of $(b_2,b_4)$
		determined within the BMS approach \cite{Bakulev:2001pa} for two different values of
		$\lambda_q^2=0.4$~GeV$^2$ (larger shaded rectangle) and 0.45~GeV$^2$ (transparent rectangle), where
		the BMS DA \cite{Bakulev:2001pa} is represented by \ding{54}.
		The smaller shaded rectangle encloses the range of $(b_2,b_4)$ coefficients associated with
		DAs having a platykurtic profile \cite{Stefanis:2015qha}, like the model \ding{60} proposed
		in \cite{Stefanis:2014nla}.
		The dash-dotted, dashed (red), and solid (blue) vertical lines show the lattice results
		for $b_2$ from \cite{Bali:2019dqc}
		for the NLO ($0.109(37)$), NNLO ($0.139(32)$),  and N$^3$LO ($0.159_{-0.027}^{+0.025}$), respectively.
	}
\end{figure*}

\begin{figure*}[htb]
		\includegraphics[width=0.45\textwidth]{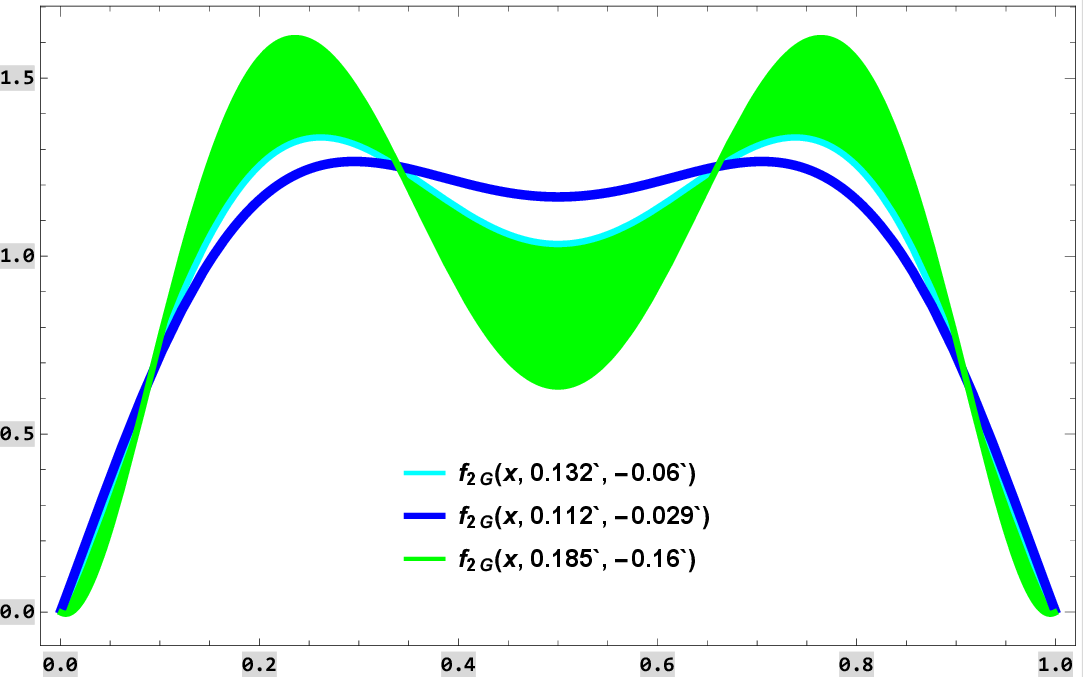}
 \vspace{-4mm}
	\caption{\label{fig:DAmodels} (color in line)
      \footnotesize
The profiles of different DAs:
a bunch of green color is the admissible domain for BMSmod DA \GrayW{\ding{115}},
	while the curve navy in color corresponds to the best fit DA \BluTn{\ding{108}}
$(b_2^{bf}=0.112, b_4^{bf}= -0.029)$ for the pion TFF,
compared with the profile of the DA obtained in the lattice approach LaMET in \cite{Baker:2024zcd}.}
\end{figure*}

 The parametric presentation $(b_2,b_4)$ for different pion DAs in Fig.\ref{fig:fitTFF-NNLO}
 is taken from  \cite{Mikhailov:2021znq};
  and some of corresponding profiles of pion DAs presented in Fig.\ref{fig:fitTFF-NNLO} are shown in Fig.\ref{fig:DAmodels}.
   Both  the BMSmod \GrayW{\ding{115}} and the platykurtic DA \ding{60} belong to the 1$\sigma$
level (red ellipse) of the best-fit point DA$^{bf}$ \BluTn{\ding{108}}.
\vspace*{-3mm}
 \vspace*{-5mm}
 \end{appendix}

\bibliographystyle{apsrev}
\newcommand{\noopsort}[1]{} \newcommand{\printfirst}[2]{#1}
  \newcommand{\singleletter}[1]{#1} \newcommand{\switchargs}[2]{#2#1}

\end{document}